\documentclass[pdflatex,sn-mathphys-num]{sn-jnl}% Math and Physical Sciences Numbered Reference Style 
\usepackage{graphicx}%
\usepackage{multirow}%
\usepackage{amsmath,amssymb,amsfonts}%
\usepackage{amsthm}%
\usepackage{mathrsfs}%
\usepackage[title]{appendix}%
\usepackage{xcolor}%
\usepackage{textcomp}%
\usepackage{manyfoot}%
\usepackage{booktabs}%
\usepackage{algorithm}%
\usepackage{algorithmicx}%
\usepackage{algpseudocode}%
\usepackage{listings}%
\usepackage{makecell}
\usepackage{subfigure}
\usepackage{svg}

\usepackage{anyfontsize}

\begin{document}

\title[Article Title]{Artificial Intelligence Driven Workflow for Accelerating Design of Novel Photosensitizers}


\author*[1]{\fnm{Hongyi} \sur{Wang}}\email{howard.wang@my.cityu.edu.hk}
\author*[2]{\fnm{Xiuli} \sur{Zheng}}\email{zhengxiuli@mail.ipc.ac.cn}
\author[2]{\fnm{Weimin} \sur{Liu}}%\email{wmliu@mail.ipc.ac.cn}
\author[3]{\fnm{Zitian} \sur{Tang}}
% \author[4]{\fnm{Tao} \sur{Wang}}
% \author[4]{\fnm{Kaiji} \sur{Sun}}
% \author[6]{\fnm{Ji} \sur{Sun}}
% \author[1]{\fnm{Li} \sur{Zhai}}
% \author[1]{\fnm{Zijian} \sur{Li}}
% \author[1]{\fnm{Wei} \sur{Zhai}}
% \author[4]{\fnm{Xusheng} \sur{Wang}}
\author*[4]{\fnm{Sheng} \sur{Gong}}\email{shenggongmit@gmail.com}
% \author*[1]{\fnm{Bolong} \sur{Huang}}\email{bolhuang@cityu.edu.hk}
% \author*[1]{\fnm{Hua} \sur{Zhang}}\email{hua.zhang@cityu.edu.hk}

\affil[1]{\orgdiv{Department of Chemistry}, \orgname{City University of Hong Kong}, \orgaddress{\city{Kowloon}, \state{Hong Kong},  \country{China}}}
\affil[2]{\orgdiv{Technical Institute of Physics and Chemistry}, \orgname{Chinese Academy of Sciences}, \orgaddress{\city{Beijing}, \country{China}}}
\affil[3]{\orgdiv{Department of Biomedical Engineering}, \orgname{City University of Hong Kong}, \orgaddress{\city{Kowloon}, \state{Hong Kong},  \country{China}}}
% \affil[4]{\orgdiv{Institute of Chemistry}, \orgname{Chinese Academy of Sciences}, \orgaddress{\city{Beijing}, \country{China}}}
% \affil[4]{\orgname{Independent Researcher}, \orgaddress{\city{Stockholm}, \country{Sweden}}}
% \affil[4]{\orgdiv{School of Mathematics}, \orgname{Renmin University of China}, \orgaddress{\state{Beijing}, \country{China}}}
% \affil[6]{\orgname{ByteDance Inc.}, \orgaddress{\state{Beijing}, \country{China}}}
\affil[4]{\orgdiv{Department of Materials Science and Engineering}, \orgname{Massachusetts Institute of Technology}, \orgaddress{\city{Boston}, \state{MA 02139}, \country{USA}}}
% \affil[7]{\orgdiv{Department of Applied Biology and Chemical Technology}, \orgname{The Hong Kong Polytechnic University}, \orgaddress{\city{Kowloon}, \state{Hong Kong}, \country{China}}}

\abstract{The abstract}

%%================================%%
%% Sample for structured abstract %%
%%================================%%

\abstract{The discovery of high-performance photosensitizers has long been hindered by the time-consuming and resource-intensive nature of traditional trial-and-error approaches. Here, we present \textbf{A}I-\textbf{A}ccelerated \textbf{P}hoto\textbf{S}ensitizer \textbf{I}nnovation (AAPSI), a closed-loop workflow that integrates expert knowledge, scaffold-based molecule generation, and Bayesian optimization to accelerate the design of novel photosensitizers. The scaffold-driven generation in AAPSI ensures structural novelty and synthetic feasibility, while the iterative AI-experiment loop accelerates the discovery of novel photosensitizers. AAPSI leverages a curated database of 102,534 photosensitizer-solvent pairs and generate 6,148 synthetically accessible candidates. These candidates are screened via graph transformers trained to predict singlet oxygen quantum yield ($\phi_\Delta$) and absorption maxima ($\lambda_{max}$), following experimental validation. This work generates several novel candidates for photodynamic therapy (PDT), among which the hypocrellin-based candidate HB4Ph exhibits exceptional performance at the Pareto frontier of high quantum yield of singlet oxygen and long absorption maxima among current photosensitizers ($\phi_\Delta$=0.85, $\lambda_{max}$=650nm). The database will be available after acceptance.}




\maketitle

\section{Introduction}\label{sec_intr}

Photodynamic therapy (PDT) has emerged as one of the most promising and rapidly advancing modalities within the field of precision medicine, owing to its precise temporal and spatial controllability, non-invasiveness, minor side effects, and low likelihood of drug resistance~\cite{zhao2025pdt11, li2025pdt12, chen2024pdt13}. It is extensively applied in the treatment of a variety of clinical conditions, including both superficial malignant and benign lesions, as well as infectious diseases~\cite{rajora2017pdt21, wang2024pdt22, zhao2021pdt23, li2025pdt24}. The technique works by activating a photosensitizer through laser irradiation, leading to the production of reactive oxygen species (ROS), which in turn induce damage and death in cancer cells. Therefore, light, oxygen, and PS are the three essential components required to achieve PDT, among which photosensitizers are the key factors in achieving PDT. However, conventional methods used in the design of photosensitizers are often encumbered by inherent challenges, including the laborious and time-intensive nature of experimental screening processes, limited exploration of chemical space, and the intricate balance required among a set of physical chemical properties such as the quantum yield of singlet oxygen, solubility, and absorption pattern~\cite{abrahamse2016PDT_review_intro, lan2019PDT_review_intro}. These challenges underscore the pressing need for innovative and efficient strategies to expedite the discovery and optimization of photosensitizers with enhanced performance characteristics. 

In response to these challenges, computational methods are incorporated to accelerate the design of potential photodynamic drug molecules. Density functional theory (DFT) and time-dependent density functional theory (TD-DFT) are the most applied \textit{ab initio} computational methods in studying photosensitizers as both ground state and excited states are involved in light absorption~\cite{drzewiecka2021tddft_review_intro}. Properties such as absorption 
spectra, singlet-triplet gap and HOMO-LUMO gap can be computed by DFT~\cite{batra2020tddft_intro, rintoul2013tddft_intro, xu2021PS_bayesian_intro}. However, DFT and TD-DFT are not efficient for large systems due to the high scaling of computational demand with respect to the system size ($\geq \mathcal{O}(N^3)$), making the \textit{ab initio} methods not suitable for rapid screening of large photosensitizer systems. 

Recently, machine learning (ML) has been extensively applied to drug discovery, with various types of tasks such as property prediction~\cite{segall2014intro_pred1, tao2015intro_pred2, cai2022midruglikeness}, molecule generation~\cite{de2018molgan, lim2020molvae, bagal2021molgpt, maziarz2021moler}, molecular dynamics~\cite{gong2024bamboo, dpa2, zhang2018dp, zhang2023dp_thermal}, synthetic route prediction~\cite{coley2018intro_route1, genheden2022fintro_route2, jiang2023intro_route3}, and automated experiment~\cite{coley2019intro_auto1, gromski2020intro_auto2, grisoni2021intro_auto3, goldman2022intro_auto4}. Among the extensive exploration of applying ML in drug discovery, most applications of ML in design of photosensitizer focus on data-driven property prediction~\cite{greenman2022peak_ML_intro, he2024PS_prediction_intro, noto2023classifier_ML_intro, buglak2021QSPR_predict_intro}. Trained with experimental and computational data, these models are able to predict the physical and chemical properties of photosensitizers with low cost. With the rapid development of artificial intelligence (AI) technologies, generative models with Bayesian optimization (BO)~\cite{shahriari2015bayesian, snoek2012bayesian, xu2021PS_bayesian_intro} and reinforcement learning~\cite{li2023PS_generative_intro_arxiv} are applied in recent works to generate new photosensitizer candidates. Furthermore, Xu et al.~\cite{xu2021PS_bayesian_intro} combines AI model with traditional TD-DFT computations to enable more efficient discovery of photosensitizers. Despite substantial progress in compiling photosensitizer datasets and applying predictive AI models for property optimization, critical gap persists. First, generative AI, particularly scaffold-based \textit{de novo} design, remains unexplored for photosensitizer discovery. The complexity of the photosensisitizers and the lacking of experiment-derived database of excited-state properties hinder the domain-specific finetuning of AI models. Second, no AI-generated photosensitizer has yet undergone experimental validation demonstrating clinically relevant photodynamic therapy potential. Current efforts predominantly focus on virtual screening of known libraries or single-property prediction, lacking closed-loop frameworks that integrate generative design, multi-objective optimization, and experimental feedback. This absence of generative approaches and experimental data represents a bottleneck in harnessing AI for accelerated PDT agent development. 
% attempt to accelerate 

\begin{figure}
    \centering
    \includegraphics[width=0.99\linewidth]{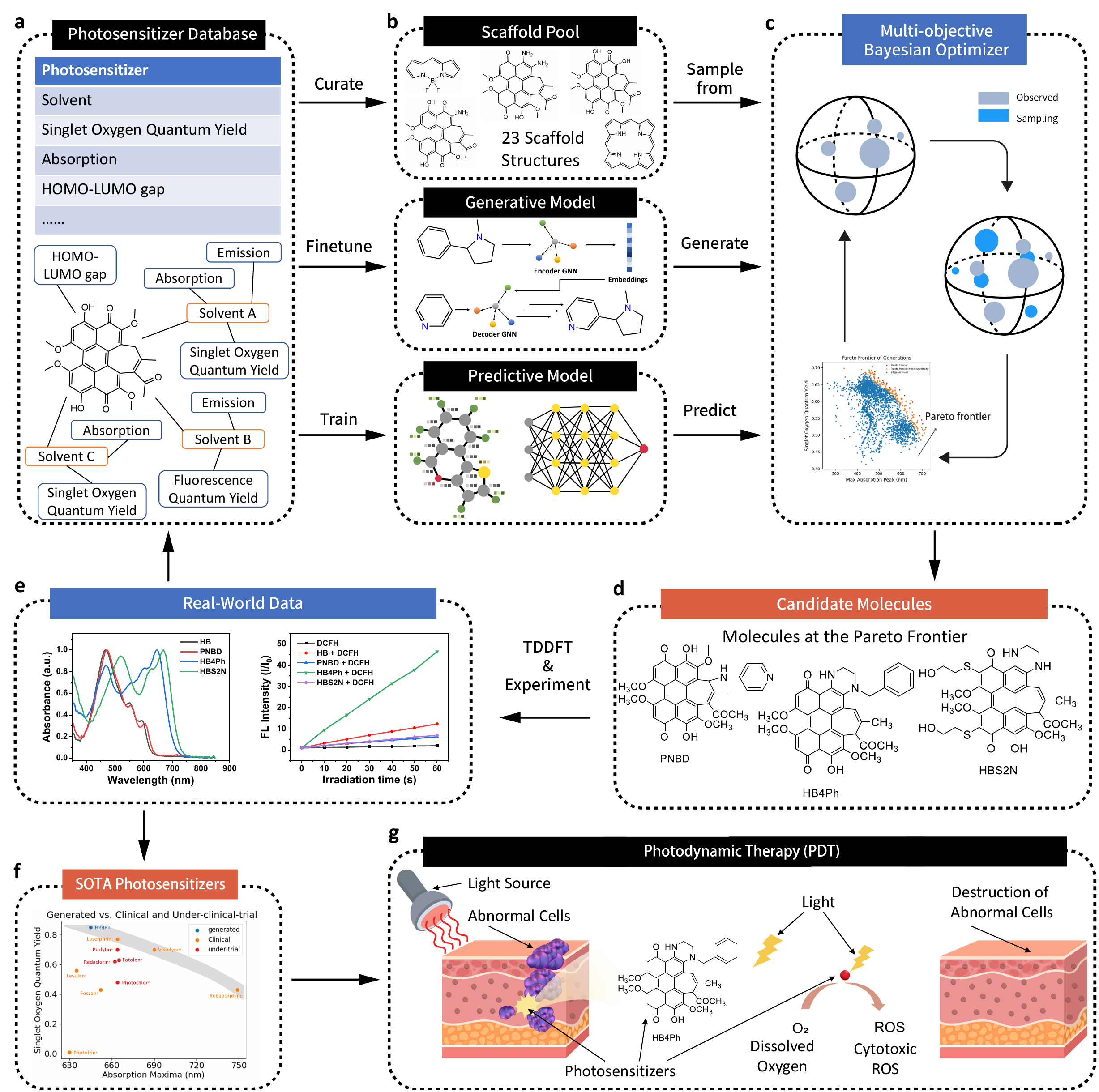}
    \caption{This figure illustrates the AAPSI workflow. \textbf{a}: We collect a photosensitizer database that includes published and unpublished data. \textbf{b}: Key elements of this workflow is derived from the database. A set of scaffold structures are selected and modified from the molecules in the database to incorporate expert knowledge and ensure the validity of the generated structures. We finetune a MoLeR model for molecule generation from a pretrained checkpoint and train a SolutionNet model for property prediction from scratch. \textbf{c}: Unified with MOBO, we use the scaffold pool, the MoLeR model, and the SolutionNet model to iteratively generate molecules. After the generation, the Pareto frontier of high $\phi_\Delta$ and long wavelength of $\lambda_{max}$ is identified. \textbf{d}: We manually screen the generations and select some synthetically accessible molecules as the candidate molecules. \textbf{e}: Three of the candidate molecules as selected for further investigations, including TD-DFT calculations, synthesis, and characteristics. Real-world data is subsequently added to the database. \textbf{f}: Among the synthesized molecules, HB4Ph emerged as the first experimentally validated, AI-designed photosensitizer demonstrating high PDT potential, with a $\phi_\Delta$ of 0.85 and an $\lambda_{max}$ at 645 nm. Compared with clinical-used and under-trial drugs, HB4Ph is at the Pareto frontier of the two properties. \textbf{g}: A demonstration of PDT, the subsequent application of the photosensitizers.}
    \label{fig:scheme}
\end{figure}

In this work, we present \textbf{A}I-\textbf{A}ccelerated \textbf{P}hoto\textbf{S}ensitizer \textbf{I}nnovation (AAPSI, pronounced as AP-see), an integrated workflow that accelerates the discovery of novel photosensitizers with experts' knowledge and AI methods, including scaffold-based molecule generation, graph transformer for property prediction, and multi-objective Bayesian optimization (MOBO), as illustrated in Figure~\ref{fig:scheme}. In addition to the AI methods, the database of photosensitizers and the curated set of scaffolds, from which new molecules were generated, induced the knowledge of experts to this workflow. This knowledge facilitates the rapid generation of novel photosensitizer candidates with tailored properties from natural products. In the molecule generation section, 6,148 molecules were generated from 23 scaffolds. These new molecules were then screened by predictive graph transformers and the molecules at the Pareto frontier were further investigated in experiments. HB4Ph, a hypocrellin-based candidate, achieves the state-of-the-art performance as photosensitizers on singlet oxygen quantum yield ($\phi_\Delta$ = 0.85) and absorption maxima ($\lambda_{max}$ = 645 nm) upon experimental validation, demonstrating superior PDT capability and confirming the efficiency of AAPSI for tailored molecular design. HB4Ph lies on the Pareto frontier of in clinical and under-trial organic molecular photosensitizers for PDT, optimally balancing a high singlet oxygen quantum yield with deep-tissue-penetrating absorption maxima. This closed-loop integration of AI, expert knowledge, and experimental validation demonstrates AAPSI’s potential to reduce discovery timelines for next-generation photosensitizers and proves the efficiency of AI-driven exploration paradigm for material innovation. We summarize our main contributions as follows:

\begin{enumerate}
    \item We establish an integrated workflow to accelerate the design of photosensitizers functioning through generation, screening and evaluation.
    \item We establish a database containing 102,534 photosensitizer-solvent pairs with target property, comprising 23,650 unique photosensitizers, covering a wide range of relevant properties. The AAPSI database is available online: http://aapsi.online.
    % \item We trained predictive graph transformers with our database for predicting singlet oxygen quantum yield and absorption maxima.
    % \item We generate 3,660 photosensitizer candidates directly using the scaffold-based generative model finetuned on our database, as well as 2,488 photosensitizer candidates with multi-objective optimized molecule generation.
    \item We generate 6,148 photosensitizers and synthesize and characterize 3 molecules at the Pareto frontier of highest $\phi_\Delta$ and longest wavelength of $\lambda_{max}$. HB4Ph derived from hypocrellin exhibits exceptional photophysical properties ($\phi_\Delta$ = 0.85, $\lambda_{max}$ = 645 nm) as the first AI-generated state-of-the-art photosensitizer for PDT.
\end{enumerate}

\subsection{The AI-driven Workflow for discovery of novel photosensitizers}\label{rslt_data}
 
\subsubsection{AI-driven Workflow}\label{rslt_wkfl}

We establish AAPSI, an AI-driven workflow (illustrated in Figure~\ref{fig:scheme}) that integrates dataset curation, scaffold-based generation, machine learning-guided virtual screening, and experimental validation to accelerate photosensitizer discovery.

In the first phase, we construct a comprehensive database of 102,534 photosensitizer-solvent pairs (23,650 unique photosensitizer), combining published and proprietary data. This dataset spans diverse structural classes and critical photodynamic properties, including $\phi_\Delta$ and $\lambda_{max}$. To consider synthetic feasibility and domain expertise, 23 scaffolds—derived from natural products such as hypocrellin—are curated for controlled molecule generation. A pretrained scaffold-based generative model (MoLeR~\cite{maziarz2021moler}) is finetuned on a dataset that merges the Guacamol~\cite{brown2019guacamol} and the AAPSI database to prioritize synthetically accessible designs, while a graph transformer model (SolutionNet) is trained to predict $\phi_\Delta$ and $\lambda_{max}$ with uncertainty quantification.

The second phase generates 6,148 novel candidates through two complementary strategies: scaffold-based generation (3,660 candidates) and MOBO (2,488 candidates). The latter balanced $\phi_\Delta$, $\lambda_{max}$, and synthetic accessibility, guided by Pareto frontier analysis. Screening with the graph transformer prioritizes 9 synthetically accessible candidates at the Pareto frontier, where $\phi_\Delta$ and $\lambda_{max}$ are maximized simultaneously.

In the final phase, three top candidates (HB4Ph, PNBD, HBS2N) are synthesized and experimentally validated. Among these, HB4Ph, a hypocrellin derivative, emerges as a promising candidate, achieving $\phi_\Delta$ = 0.85 (exceeding most clinical photosensitizers) and $\lambda_{max}$ = 645 nm, ideal for deep-tissue PDT. Its structure-activity relationship aligns with predictions from SolutionNet, confirming the workflow’s ability to link molecular features to tailored photodynamic properties.

\subsubsection{Database of Photosensitizers}

\begin{figure}[ht]
    \centering
    \includegraphics[width=0.99\linewidth]{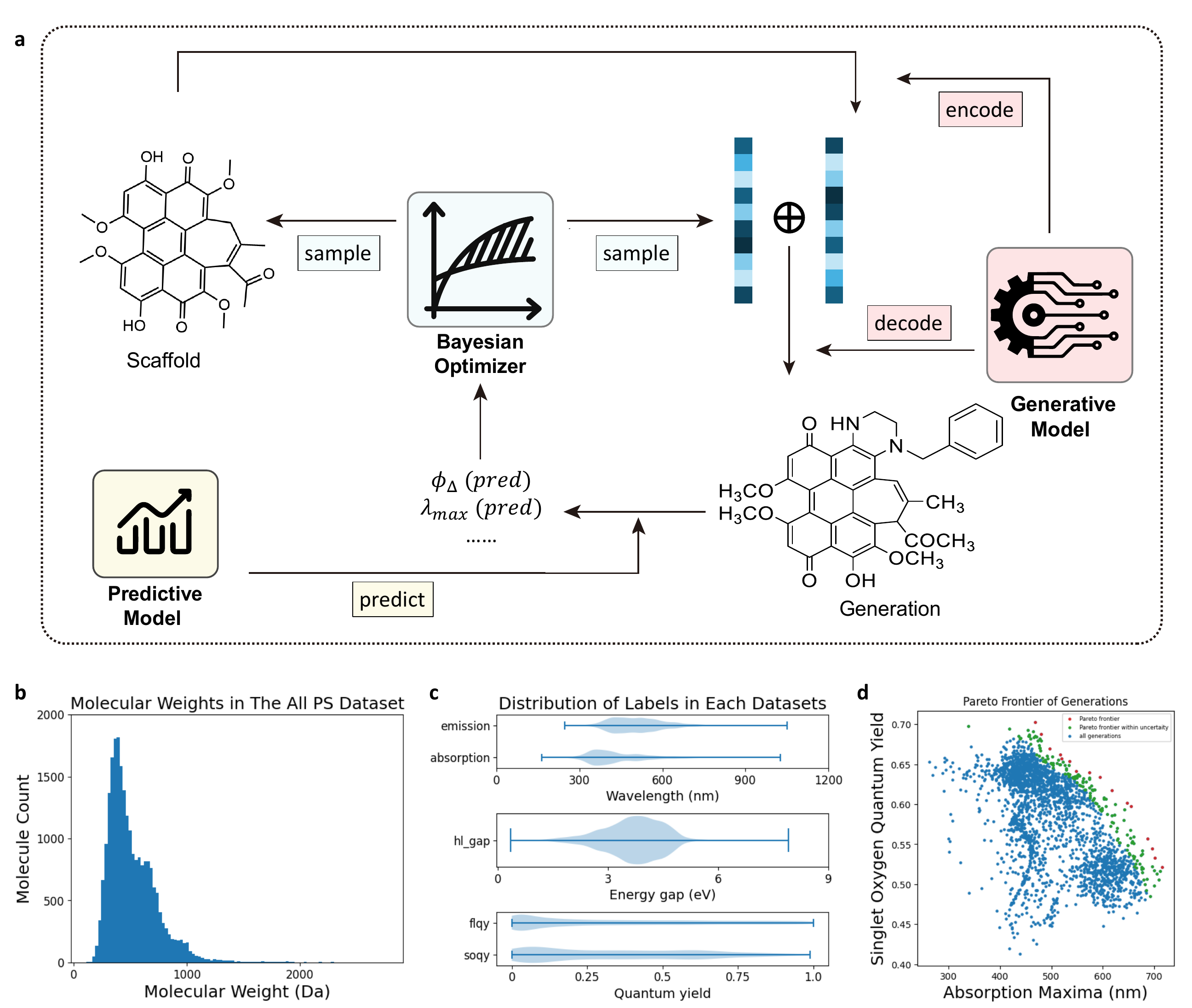}
    \caption{\textbf{a}: An overview of the generation process. The Bayesian optimizer samples a scaffold and a hidden bias representation. The generative model encodes the scaffold into another hidden representation, which is added with the bias representation, then decodes the hidden representation into a new molecule. The predictive model screens the new molecule and predicts the target properties, which is applied to the posterior probability calculation in the next cycle. \textbf{b}: Statistics of dataset \texttt{all\_ps} on molecular weight. \textbf{c}: Range and distribution of labels of other datasets. \textbf{d}: Demonstration of generated results from direct generation and Bayesian multi-objective optimization, with the Pareto frontier of $\phi_\Delta$ and $\lambda_{max}$ highlighted.}
    \label{fig_ps:database}
\end{figure}

This photosensitizer database comprises 23,650 unique photosensitizers and 102,534 photosensitizer-solvent pairs associated with specific target properties. The photochemical properties of photosensitizers depend on the solvent, as solvents mediate key solute-solvent interactions that directly alter the molecule’s electronic structure and excited-state decay pathways, thereby modifying solvent-sensitive properties. Thus, besides datasets that either serve as a repositories without designated target properties (\texttt{all\_ps} dataset) or focus on intrinsic molecular properties unaffected by solvent environments (\texttt{hl\_gap} dataset), all other datasets in the database adopt photosensitizer-solvent pairs as the core input. This design ensures that solvent-induced variations in target photochemical properties can be adequately captured, aligning with the experimental reality of photosensitizer performance in solution-based applications. The molecular weights of the photosensitizers range from 110 to 2,779 Dalton, with the majority falling below 750 Dalton, as illustrated in the statistics in Figure~\ref{fig_ps:database}a. The target properties also cover a wide range with distributions illustrated in Figure~\ref{fig_ps:database}b. 

This database contains both open-source data compiled from existing databases and literature, as well as proprietary data not published elsewhere. The molecules in the database are structurally classified into derivatives of porphyrin, BODIPY, phenothiazine, xanthene, cyanine, phthalocyanine, chlorin, and perylenequinone. A significant portion of the unpublished data falls under the perylenequinone category, including hypocrellin derivatives, which are regarded as next-generation photosensitizers for PDT. 

The database is organized into six subsets, each focusing on a different target to facilitate a wide range of applications such as PDT, bioimaging, and optoelectronics. These subsets include absorption maxima, emission maxima, HOMO-LUMO gaps, singlet oxygen quantum yield, fluorescence quantum yield, and a collection of all photosensitizers. Detailed information concerning the database is presented in Table~\ref{tab_ps:database_01}.

\begin{table}[]
    \tiny
    % \resizebox{0.9\textwidth}{!}{
    \centering
    \caption{Details regarding the photosensitizer database are summarized in this table, including database size, descriptions of target labels, and the corresponding ranges of these labels. The \texttt{all\_ps} dataset comprises all photosensitizer entries within this database, serving as a comprehensive repository of photosensitizer-related information. The \texttt{absorption} and \texttt{emission} datasets utilize photosensitizer-solvent pairs as input, with the wavelength of absorption maxima and emission maxima of the dissolved photosensitizer designated as their respective target properties. The \texttt{hl\_gap} dataset associates each photosensitizer with its corresponding HOMO-LUMO gap. The \texttt{soqy} and \texttt{flqy} datasets take photosensitizer-solvent pairs as input features, with the singlet oxygen quantum yield and fluorescence quantum yield of the photosensitizer in the specified solvent defined as their target properties, respectively.}
    
    % \resizebox{0.9\textwidth}{!}{
    \begin{tabular}{c|cccccc}
    \hline
         Dataset name&   \texttt{all\_ps} & \texttt{absorption} & 
 \texttt{emission} & \texttt{hl\_gap} & \texttt{soqy} & \texttt{flqy} \\
 \hline
 Size & 23,763& 22, 752& 23,437& 14,213& 954&17,716\\
 Labels& N/A& \makecell[c]{Wavelength of \\absorption \\ maxima (nm)} & \makecell[c]{Wavelength of \\emission \\ maxima (nm)}&  \makecell[c]{HOMO-LUMO \\gap (eV)}& \makecell[c]{Singlet oxygen \\quantum yield \\ (unitless)}& \makecell[c]{fluorescence \\quantum yield \\ (unitless)}\\
 Minimum value& N/A& 162.0& 247.0& 0.35& 0.00&0.00\\
 Maximum value& N/A& 1026.0& 1050.0& 7.91& 0.99&1.00\\
 \hline
 
 \end{tabular}
 % }
    \label{tab_ps:database_01}
\end{table}

\subsubsection{Molecule Generation and Multi-objective Optimization}\label{rslt_bay_mol}

To address the challenges in the design of photosensitizers, we focus on the optimization of the two key properties, absorption maxima and singlet oxygen quantum yield. This abstraction is rooted in the core mechanism and clinical demands of PDT, the key application of photosensitizers. First, singlet oxygen ($^1O_2$) is the primary reactive oxygen species (ROS) mediating PDT’s cytotoxicity. A higher $\phi_\Delta$ means the photosensitizer generates more $^1O_2$ upon light activation, directly enhancing its ability to kill abnormal cells and improving therapeutic efficacy. Second, $\lambda_{max}$ determines the tissue penetration depth of activating light: longer wavelengths minimize absorption and scattering by biological tissues, enabling treatment of deeper sites while reducing damage to superficial tissues. Additionally, these two properties are often mutually constrained in traditional design, making them the most critical, clinically relevant objectives to prioritize for optimizing PDT performance. In this work, we do not consider water solubility, synthetic accessibility, and other properties as primary properties because the core goal is to develop PDT, where $\phi_\Delta$ and $\lambda_{max}$  are the two most critical properties that fundamentally determine PDT efficacy and clinical applicability, while other properties, though important for practical application, are secondary attributes that can be effectively screened and evaluated after generating molecules without compromising the prioritization of the core PDT-related performance metrics.

We conducted two different molecule generation processes, direct scaffold-based molecule generation and themolecule generation with multi-objective optimization of noise embedding and scaffold selection. In the direct molecule generation, we produced 500 molecules for each scaffold, resulting in a total of 11,500 molecules, of which 3,660 are unique. In the multi-objective optimized generation, we generated 3,200 molecules sampled from the posterior probabilistic distribution of the Pareto frontier, yielding 2,488 unique molecules. The molecules generated directly exhibit less diversity, as they closely resemble the scaffolds in both structure and physical properties. In contrast, the molecules produced through multi-objective Bayesian optimization demonstrate greater diversity, facilitating the discovery of compounds with higher $\phi_\Delta$ for higher therapeutic efficacy and a longer $\lambda_{max}$ for deeper tissue penetration in medical uses. Additionally, the molecules generated via Bayesian sampling show slightly improved water solubility, although they tend to be more difficult to synthesize.

% \subsubsection{Pareto Frontier of Multiple Objectives}\label{rslt_bay_mol}

We show the Pareto frontier of the longest $\lambda_{max}$ and the highest $\phi_\Delta$, with the generated molecules plotted in Figure~\ref{fig_ps:database}c. Molecules generated from multi-objective Bayesian optimization push the Pareto frontier outward iteratively. 

Among the generated molecules at the Pareto frontier, we manually selected 9 potential photosensitizer molecules that are synthetically accessible and demonstrated them in Figure~\ref{fig_ps:molecules} and Figure~\ref{fig_ps:si_other_molecules}, with predicted properties. Some selected molecules exhibit high $\phi_\Delta$, some exhibit long $\lambda_{max}$, and some are superior in both properties. These molecules are selected from the generated molecules to ensure the stability and validity of the potential drug by manually removing molecules with unstable or inappropriate functional groups such as peroxides or very bulky alkyl groups. Although we did not dive into all the potential photosensitizers at the frontier, we believe that they may represent promising candidates for PDT. 

\subsection{Novel Photosensitizers}\label{rslt_exp}

\begin{figure}
    \centering
    \includegraphics[width=0.99\linewidth]{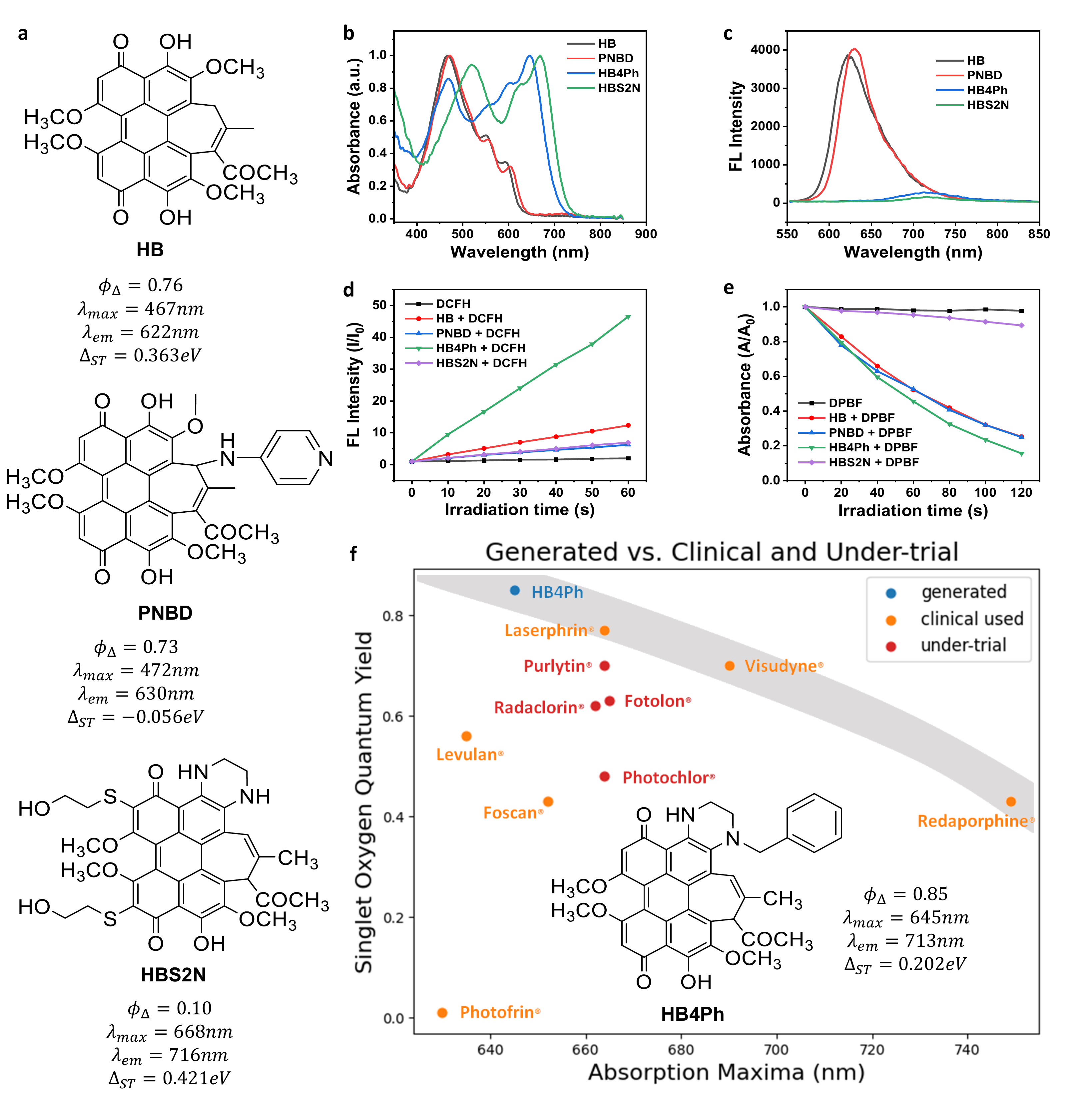}
    \caption{\textbf{a}: Molecule structure and PDT related physical properties of HB, PNBD, HBS2N, and HB4Ph (in the middle of subfigure f). Some characterization of the four molecules are illustrated in this figure, including \textbf{b}: absorption spectra; \textbf{c}: emission spectra; \textbf{d}: ROS detection; \textbf{e}: singlet oxygen detection. \textbf{f}: The comparison of HB4Ph with clinical-used and under-trial photosensitizers for PDT. HB4Ph emerges at the Pareto frontier of $\lambda_{max}$ and $\phi_\Delta$.}
    \label{fig_ps:molecules}
\end{figure}

The selection of the target molecules follows a multi-step, rigorous screening process. First, molecules are generated via MOBO generation. These candidates are screened using SolutionNet to predict $\phi_\Delta$ and $\lambda_{max}$ during the MOBO generation, followed by Pareto frontier analysis. Only molecules lying close to the frontier are retained, as they represent optimal trade-offs between the two key properties. The search space of potential photosensitizers is reduced to the synthetically accessible molecules at the Pareto frontier, which is possible to be assessed by human experts. Next, synthetic feasibility is evaluated: molecules with unstable functional groups, such as multiple peroxides or overly bulky alkyl groups, are excluded. The other molecules are evaluated based on the SA scores~\cite{ertl2009rdkit_sas} and experience, narrowing down to 9 potential candidates. Among these, three hypocrellin derivatives are selected for further study, as hypocrellin is a natural product-derived scaffold with proven photodynamic potential. Some other generated molecules at the Pareto frontier are listed in Figure~\ref{fig_ps:si_other_molecules}. The three selected molecules (PNBD, HB4Ph, HBS2N) are presented in Figure~\ref{fig_ps:molecules} along with hypocrellin B (HB) as the baseline. We perform time-dependent density functional theory (TD-DFT) computations on the three candidate molecules for the energy gap ($\Delta$E) between the first singlet excited state (E$_{S1}$) and the first triplet excited state (E$_{T1}$), synthesize them, and then test the absorption spectra, fluorescence spectra, ROS generation, and $\phi_\Delta$ of the three molecules. The characterization results are shown in Figure~\ref{fig_ps:molecules}b-e. Details of synthesis and characterization can be found in the Supplementary Information~\ref{si_exp}. 

% properties: abs peak, fl peak, soqy (how is it calculated?)

% Among the three selected molecules, HB4Ph exhibits the greatest potential as a photosensitizer candidate with the maximum absorption peak at 645 nm and a singlet oxygen quantum yield of 0.85. Although the other two photosensitizer candidates are not as compatible as HB4Ph, they also show the potential photosensitizer for PDT to some extent.  

The photophysical properties of three synthesized photosensitizers—HB4Ph, HBS2N, and PNBD—were characterized and compared to the natural product HB. Key parameters, including absorption/fluorescence profiles, fluorescence intensity, and $\phi_\Delta$, were analyzed to evaluate their potential as photodynamic therapy (PDT) agents. Structural features were correlated with their performance to elucidate design principles. The structure and photophysical characterizations are shown in Figure~\ref{fig_ps:molecules}.

HB4Ph exhibits a significant red shift in both absorption and emission spectra ($\lambda_{max}$ = 645 nm and $\lambda_{em}$ = 713 nm) compared to HB, aclinically relevant advancement for deep-tissur tumor PDT. The weak fluorescence also indicates efficient intersystem crossing (ISC) to the triplet state. This is consistent with its remarkably high $\phi_\Delta$ (0.85), higher than all clinical-used drugs, at the state-of-the-art level of all photosensitizers. The integration of two nitrogen atoms into its conjugated $\pi$-system likely enhances spin-orbit coupling, promoting ISC and subsequent singlet oxygen generation. The extended conjugation also shifts absorption close to the near-infrared (NIR) therapeutic window (700–1200 nm), enabling deeper tissue penetration—a critical advantage for clinical PDT.

HBS2N, featuring two nitrogen and two sulfur atoms in its conjugated system, displays the longest wavelength of $\lambda_{max}$ (668 nm) but the lowest $\phi_\Delta$ (0.1). Despite the heavy-atom effect from sulfur, which theoretically promotes ISC, the low singlet oxygen yield suggests competing non-radiative decay pathways or destabilization of the triplet state. Sulfur’s larger atomic radius and polarizable electron cloud may disrupt the optimal geometry for energy transfer to molecular oxygen, highlighting a trade-off between red-shifted absorption and photodynamic efficacy.

PNBD contains a nitrogen and pyridine group in non-conjugated regions, shows absorption and fluorescence profiles similar to HB (absorption: 472 nm vs. 467 nm; emission: 630 nm vs. 622 nm). Its strong fluorescence correlates with a moderately high $\phi_\Delta$ (0.73), slightly lower than HB ($\phi_\Delta$ = 0.76). The absence of conjugated heteroatoms limits red-shifting, confining its absorption to relatively short wavelength, which restricts utility in deep-tissue applications. However, the non-conjugated pyridine moiety may enhance solubility or target affinity without perturbing the core photophysical mechanism.

HB4Ph emerges as the most promising candidate, combining NIR absorption, minimal fluorescence competition (indicative of efficient triplet-state population), and unparalleled singlet oxygen generation. Its $\phi_\Delta$ of 0.85 surpasses clinical benchmarks for PDT agents, suggesting potent cytotoxicity even at low irradiation doses. While HBS2N’s absorption is deeper in the NIR range, its negligible singlet oxygen production renders it ineffective. PNBD, though comparable to HB in $\phi_\Delta$, lacks the spectral advantages of NIR light. The strategic placement of nitrogen atoms in HB4Ph’s conjugated system exemplifies how targeted structural design can optimize both light absorption and ROS generation, addressing two major challenges in PDT: tissue penetration and therapeutic efficiency.

\subsection{Database and Scaffolds}\label{meth_database_scaffold}

\subsubsection{Database of Photosensitizers}

The database utilized in this study comprises a collection of photosensitizer-solvent pairs, encompassing 23,763 unique photosensitive molecules. This dataset was meticulously curated through a combination of manual extraction from published literature and integration from existing databases~\cite{ju2021chemfluor_data, joung2020db_chromophore_data, xu2021PS_bayesian_intro}.

\begin{figure}
    \centering
    \includegraphics[width=0.9\linewidth]{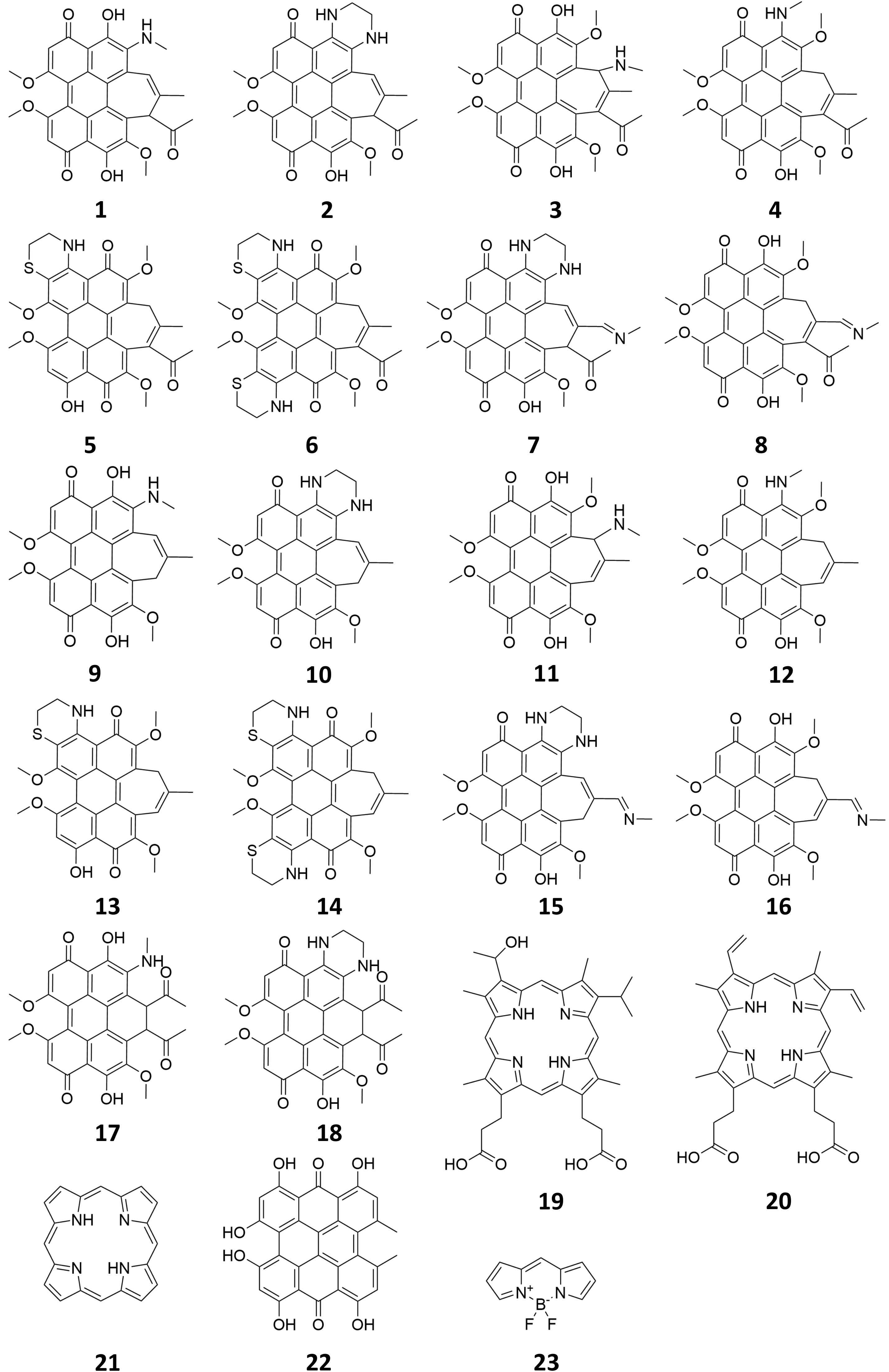}
    \caption{The scaffold where molecule generation start from. Molecule 1 represents HB the natural product, molecule 2 represents HB4 in section~\ref{syn_hb4}.}
    \label{fig_ps:scaffold}
\end{figure}

To facilitate targeted investigations, the database is divided into six distinct subsets, each focusing on specific molecular properties of interest. Each subset contains sample-target pairs, except for the all-molecules subset, which exclusively includes photosensitizers.

The representation of molecules within the database adheres to the simplified molecular-input line-entry system (SMILES) format, ensuring compatibility and ease of analysis across various computational techniques. Notably, the HOMO-LUMO gap subset provides only photosensitive molecule as the sample, while the other subsets contain photosensitizer-solvent pairs.

\subsection{Scaffolds for Molecule Generation}

In this work, each molecule generation starts with a scaffold, and iteratively adds substitution groups to the molecule. We manually select 23 scaffolds derived from the natural products, encompassing various categories of photosensitizers, including porphyrin, BODIPY and perylenequinones. Among the perylenequinones, we collect hypocrellin, elsinochrome, hypericin, as well as a set of derivatives with heteroatoms on specific sites that have demonstrated competitive photodynamic properties as drug candidates. The 23 scaffolds are shown in Figure~\ref{fig_ps:scaffold}.

\subsection{Models and Training}\label{meth_model_training}

\subsubsection{Predictive Model and Training}\label{meth_model_training_pred}

\begin{figure}
    \centering
    \includegraphics[width=0.99\linewidth]{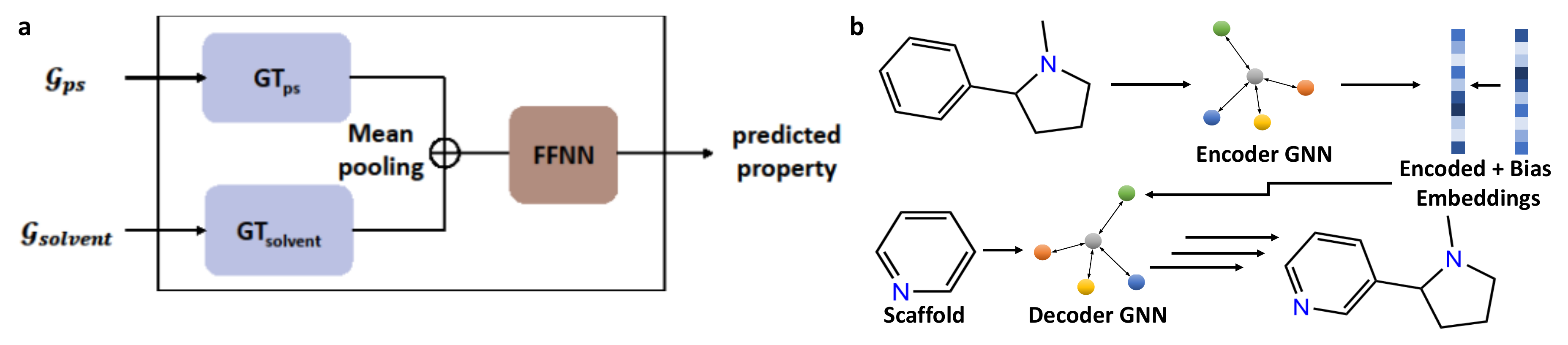}
    \caption{\textbf{a}: Overall structure of SolutionNet. The SolutionNet consists of two GT blocks and a FFNN block. The Molecule graph of photosensitizers ($\mathcal{G}_{ps}$) and solvents ($\mathcal{G}_{solv}$) are taken as input, the target property is predicted as output. Atom (node) features in $\mathcal{G}$ are encoded as $x_i$ for node $i$, while bond (edge) features between nodes $j$ and $i$ are denoted as $e_{ji}$. Positional embeddings, capturing spatial or topological information of atom $i$, are represented as $p_i$. This framework integrates structural and relational features across both graphs to model properties of photosensitizer-solvent pairs. \textbf{b}: Illustration of the structure of the generative model in this work. The framework is adopted from MoLeR~\cite{maziarz2021moler}, along with the checkpoint of the encoder-decoder. During the Bayesian optimization, the scaffold and the bias embedding is sampled by the optimizer before feeding to the decoder.}
    \label{fig:gnn_overall}
\end{figure}

The graph transformer architecture, namely SolutionNet, employed in this study is composed of two integral components, mirroring the partial structure of CrysToGraph~\cite{wang2024crystograph}. These components include the graph-wise transformer layers and feed-forward linear layers. The overall architecture is demonstrated in Figure~\ref{fig:gnn_overall}. 

The SolutionNet model can be defined as:

\begin{equation}
% \begin{center}
    SolutionNet(\mathcal{G}_{ps}, \mathcal{G}_{solvent}) = FFNN(GT_{\times N}^a(\mathcal{G}_{ps}) \oplus (GT_{\times N}^b\mathcal{G}_{solvent})) \label{eqo1}
% \end{center}    
\end{equation}

The structure of graph transformer (GT) blocks is illustrated in Figure~\ref{fig_ps:si_pred_gt}, while the feed-forward neural network (FFNN) is a naive 2-layered fully-connected layers. This graph transformer is naturally fully connected within the graph, thus, it is efficient to capture all interactions in the molecule. Details of graph construction and model structure can be found in the Supplementary Information~\ref{si_pred}.

Two independent models are trained with the subset \texttt{abs} and \texttt{soqy} for further prediction of $\lambda_{max}$ and $\phi_\Delta$ until convergence, with Adam optimizer and a learning rate of $1\times10^{-4}$. The depth of SolutionNet in this work is 3 layers for each part. The training is conducted in 5-fold cross-validation, and the one with the lowest validation error is selected for further prediction.

\subsubsection{Generative Model, Training and Generation}\label{meth_model_training_gen}

We employ a scaffold-based MoLeR model~\cite{maziarz2021moler} to generate new molecules derived from natural products.

In this study, we finetune a pretrained MoLeR checkpoint using a fine-tuning dataset, which comprising a combination of the Guacamol~\cite{brown2019guacamol} dataset and our \texttt{all\_ps} subset. The training and validation sets are randomly split into a 9:1 ratio.

Following the finetuning process, we select a set of natural products and their derivatives to serve as the scaffolds, which include perylenequinones, porphyrins, and BODIPY. Details of these scaffolds can be found in the Figure~\ref{fig_ps:scaffold}.

In the scaffold-based generation process, we first encode the scaffold molecule into a representation vector using the encoder. This representation vector is then perturbed with random noise. The decoder subsequently takes the encoded molecule as the initial scaffold and decodes the perturbed representation vector to produce a new molecule.

\subsubsection{Multi-objective Bayesian Optimization}

In the MOBO generation, we leverage the q-Noisy Expected Hyper-volume Improvement (qNEHVI)~\cite{daulton2021qnehvi} as the acquisition function over the posterior measured by a Gaussian Process surrogate model~\cite{hvarfner2024vanilla}, implemented via the ax-platform \cite{bakshy2018ae} and BoTorch~\cite{balandat2020botorch}, to efficiently sample around the Pareto frontier of multiple objectives. Our primary focus is to generate photosensitizers around the Pareto frontier, prioritizing high singlet oxygen quantum yield and long wavelength of absorption maxima in aqueous solution as the critical properties.

In the generation process, a bias vector is sampled along with the scaffold by the Bayesian optimizer. The dimension of the bias vector is then aligned to the MoLeR model with an auto-decoder. The scaffold is encoded by the finetuned MoLeR model before being added to the bias vector. A random noise is also added as a perturbation. This biased vector is finally decoded with the finetuned MoLeR model with the selected scaffold molecule as the initial structure. After the Bayesian search, the Pareto frontier is screened among all the molecules generated with biased vector and random noise.

We conducted the optimization with maximizing the two target objectives: singlet oxygen quantum yield and wavelength of absorption maxima. To predict the target properties and to quantify uncertainty, we employ the ensemble SolutionNet model trained on our datasets, with uncertainty estimated as the standard deviation of ensemble predictions. All predictions are considered in dimethylformide (DMF) solution.

\subsection{Synthesis}\label{meth_synthesis_chara} % and Characterization

\subsubsection{Synthesis of PNBD}\label{syn_pnbd}

620 mg of HB-2BF and 1 g of 4-aminopyridine were dissolved in 60 mL of tetrahydrofuran, followed by the addition of 1.5 ml of triethylamine. The mixture was heated at 55$^\circ$C for 8 h. The solvent was removed by rotary evaporation, and 30 ml of ethanol was added. The mixture was refluxed for 2 h and the solvent was removed by rotary evaporation. Impurities were removed by extraction with dichloromethane and water, and the organic layer was dried and purified by column chromatography. $^1$H NMR (600 MHz, Methanol-$d_4$) $\delta$ 7.50 (d, $J$ = 7.1 Hz, 2H), 6.99 (s, 1H), 6.58 (s, 1H), 6.54 (s, 1H), 6.38 (d, $J$ = 7.2 Hz, 2H), 4.28 (d, $J$ = 1.3 Hz, 3H), 4.10 (s, 3H), 4.08 (s, 3H), 4.05 (d, $J$ = 1.4 Hz, 3H), 2.53 (s, 3H), 2.00 (s, 3H). The NMR spectra can be found in Supplementary Information~\ref{si_exp_synth}.

\subsubsection{Synthesis of HB4}\label{syn_hb4}

HB (300 mg) and ethanediamine (2 mL) were dissolved in tetrahydrofuran (10 mL). The resultant solution was subsequently stirred in the dark at 55 $^\circ$C for 8 h. Afterward, the solvent was evaporated under reduced pressure. The residual material was suspended in dichloromethane, washed with water, and the dichloromethane was thoroughly removed. The resulting black solid was purified by silica gel column chromatography to yield the product as a dark solid. $^1$H NMR (600 MHz, Chloroform-$d$) $\delta$ 16.91 (s, 1H), 11.92 (s, 1H), 6.42 (s, 1H), 6.32 (s, 1H), 6.18 (s, 1H), 5.20 (s, 1H), 5.16 (s, 1H), 4.15 (s, 3H), 4.02 (s, 3H), 3.96 (s, 3H), 3.80 (qd, $J$ = 11.6, 9.2, 5.0 Hz, 2H), 3.76 – 3.71 (m, 1H), 3.64 (dt, $J$ = 11.3, 5.2 Hz, 1H), 2.30 (s, 3H), 1.60 (s, 3H). The NMR spectra can be found in Supplementary Information~\ref{si_exp_synth}.

\subsubsection{Synthesis of HB4Ph}\label{syn_hb4ph}

Dissolve HB4 (500 mg) and cesium carbonate (820 mg) in 150 mL of DMF, followed by stirring at room temperature for 20 minutes. Subsequently, add methyl 4-bromomethylbenzoate (580 mg) to the reaction mixture and allow the reaction to proceed at room temperature for 8 hours. Upon completion of the reaction, extract the mixture three times using an ethyl acetate/water system, and subsequently evaporate the organic phase to dryness. The resultant black solid was further purified via silica gel column chromatography, yielding the desired product as a dark solid. $^1$H NMR (400 MHz, Chloroform-$d$) $\delta$ 11.99 (s, 1H), 7.55 (d, $J$ = 7.1 Hz, 2H), 7.43 (dd, $J$ = 8.4, 6.8 Hz, 2H), 7.37 – 7.31 (m, 1H), 6.83 – 6.79 (m, 1H), 6.41 (d, $J$ = 1.8 Hz, 2H), 5.30 (s, 1H), 5.22 (d, $J$ = 1.5 Hz, 1H), 5.14 (d, $J$ = 15.9 Hz, 1H), 4.42 (d, $J$ = 15.9 Hz, 1H), 4.19 (s, 3H), 4.02 (s, 3H), 3.98 (s, 3H), 3.65 (td, $J$ = 12.5, 11.7, 3.6 Hz, 1H), 3.56 (d, $J$ = 13.4 Hz, 1H), 3.35 (dd, $J$ = 14.1, 3.1 Hz, 1H), 3.18 – 3.08 (m, 1H), 2.01 (d, $J$ = 1.5 Hz, 3H), 1.42 (s, 3H). The NMR spectra can be found in Suppementary Information~\ref{si_exp_synth}.

\subsubsection{Synthesis of HBS2N}\label{syn_hbs2n}

HB4 (300 mg) and 2-mercaptoethanol (1 mL) were dissolved in DMSO/triethylamine (20 mL/12 mL), and at room temperature for 1 h. Afterward, the solvent was evaporated under reduced pressure. The residual material was suspended in dichloromethane, washed with water, and the dichloromethane was thoroughly removed. The resulting black solid was purified by silica gel column chromatography to yield the product as a dark solid. $^1$H NMR (400 MHz, Chloroform-$d$) $\delta$ 17.61 (s, 1H), 10.93 (s, 1H), 7.18 (s, 1H), 4.37 (d, $J$ = 37.0 Hz, 2H), 4.07 (d, $J$ = 6.4 Hz, 2H), 3.97 (s, 3H), 3.86 (s, 1H), 3.78 (d, $J$ = 11.0 Hz, 3H), 3.74 (s, 3H), 3.58 (s, 3H), 3.40 – 3.03 (m, 5H), 2.71 (s, 3H), 2.46 (s, 3H). The NMR spectra can be found in Supplementary Information~\ref{si_exp_synth}.

\section{Discussion}\label{sec_disc}

The development of AAPSI represents a paradigm shift in photosensitizer design, merging AI-driven innovation with domain expertise to address the inefficiencies of traditional discovery pipelines. The experimental validation of HB4Ph, a hypocrellin-derived candidate with a high singlet oxygen quantum yield ($\phi_\Delta$=0.85) and long absorption maxima wavelength ($\lambda_{max}$=645nm), underscores the workflow’s capacity to deliver high-performance molecules tailored for photodynamic therapy (PDT). This success highlights the broader potential of integrating scaffold-based generation, predictive modeling, and multi-objective optimization to accelerate molecular discovery. AAPSI exemplifies the power of AI-human collaboration in molecular discovery, offering a closed-loop framework that bridges computational design, predictive modeling, and experimental validation. AAPSI generates 6,148 candidates from 23 natural product-derived scaffolds and identifies HB4Ph as the first AI-generated PDT-optimized photosensitizer. This work demonstrates how AI can accelerate the design of high-performance photosensitizers while maintaining synthetic tractability. The integration of expert-curated scaffolds, graph transformer predictions, and MOBO ensures that candidates are both innovative and functionally tailored. We also make this database available to public at http://aapsi.online.

The reliability of AI-generated molecules is a cornerstone of AAPSI’s efficacy. The workflow addresses reliability through scaffold constraints, which anchor molecular generation to chemically validated frameworks, ensuring structural stability and synthetic feasibility. The graph transformer’s predictive accuracy further refines screening, though its performance hinges on the diversity and quality of the training data. While the current model demonstrates strong validation metrics, its generalizability to entirely novel scaffolds may require iterative refinement as the database grows. Future efforts should also focus on expanding the chemical diversity within the AAPSI framework. While the current database is extensive, it underrepresents certain photosensitizer classes, such as transition metal complexes. Incorporating computational simulations and high-throughput experimental data could fill these gaps, enabling the discovery of motifs with unique photophysical properties. Dynamic integration of real-time experimental results into the AI training loop would also enhance predictive accuracy, creating a self-improving system that adapts to emerging data.

The current workflow prioritizes two key photophysical properties, but real-world applications to make a successful drug demand a more nuanced balance of functionality and practicality. For instance, synthetic accessibility, ADMET properties and cost. These could also benefit from further AI-human collaborations. Additionally, scaffold-based generation inherently limits exploration of uncharted chemical spaces. Hybrid approaches, combining scaffold constraints with fragment-based de novo design, might resolve this trade-off, fostering innovation while maintaining feasibility.

By streamlining the path from virtual design to experimental validation, AAPSI challenges traditional paradigms of drug discovery, offering a blueprint for accelerated innovation across materials science and biotechnology. This new paradigm is also transferrable to other applications such as discovery of other drugs or catalysts. Its emphasis on synthetic accessibility and expert integration ensures that AI-generated candidates are not only high-performing but also practical, addressing a longstanding gap in AI and experimental chemistry.

\backmatter

\section*{Declarations}

\subsection*{Conflict of Interest}

The authors declare no conflicts of interest.

% \subsection{Data Availability}

% The database is now online at .

\subsection*{Code Availability}

The source code will be available after acceptance.

\bibliography{sn-bibliography}

\noindent

\begin{appendices}

% data collection
% predictive model
% training
% generation

% synthesis
% characteristics
\newpage
\setcounter{page}{1}

\renewcommand{\thefigure}{A\arabic{figure}}
\setcounter{figure}{0}
\renewcommand{\thetable}{A\arabic{table}}
\setcounter{table}{0}
\renewcommand{\theequation}{A\arabic{equation}}
\setcounter{equation}{0}
\renewcommand{\thesection}{A\arabic{section}}
\setcounter{section}{0}

% \begin{refsection}

\section*{Artificial Intelligence Driven Workflow for Accelerating Design of Novel Photosensitizers}
\section*{Supplementary Information}\label{3:si}

\section{Details of Prediction}\label{si_pred}

\subsection{Construction of Molecule Graph}\label{si_pred_graph}

In the graph transformer framework employed in this study, atoms are meticulously represented as individual nodes, while the adjacent atoms are elegantly captured as neighboring nodes within the graphs. Neighbors are identified in a k-nearest-neighbor (k-NN) manner as the k selected as 12. This approach ensures that the intricate interplay between atoms and their interactions are effectively encapsulated for subsequent computational analysis. For each atom, number of neighbors are  generally greater than number of chemical bonds. The redundancy in edge identification is reported to positively affect the performance of models~\cite{gong2024bamboo}.

To initialize the representation process, the atom and edge embedding adhere to the CGCNN~\cite{xie2018cgcnn} standard, encompassing a comprehensive set of features. These features play a pivotal role in capturing the nuanced characteristics of the molecular structures, thereby laying the foundation for subsequent graph-based computations within the graph transformer framework. 

Our research applied a graph transformer architecture instead of the massage-passing neural networks~\cite{scarselli2008gnn} to process the graph-like data, thereby making positional encoding essential. Drawing inspiration from prior studies, this architecture integrates positional encoding to effectively handle connective and spatial information.

To explicitly model the structural and connectivity information, we adopted a multifaceted approach to positional encoding, incorporating various sources of positional data, including:

\begin{enumerate}
    \item Laplacian positional encoding~\cite{dwivedi2020lappe}: This encoding is based on the Laplacian operator, which captures the structural relationships within the graph.
    \item Random walk positional encoding~\cite{dwivedi2021rwpe}: This encoding is derived from random walk processes, providing additional insight into connectivity.
\end{enumerate}

By concatenating these positional encoding, we aim to represent both spatial and connectivity information within the graph. This comprehensive approach ensures that positional encoding capture both the absolute and relational aspects of node positions and connectivity, thereby enhancing the efficacy of the graph transformers in our model.

\subsection{Architecture of the Prediction Model}\label{si_pred_model}

The SolutionNet in this work for the property prediction task consists of two GT blocks and a FFNN block. The overall architecture can be found in Figure~\ref{fig:gnn_overall}. In this section, we introduce the details of the GT blocks.

\begin{figure}
    \centering
    \includegraphics[width=0.8\linewidth]{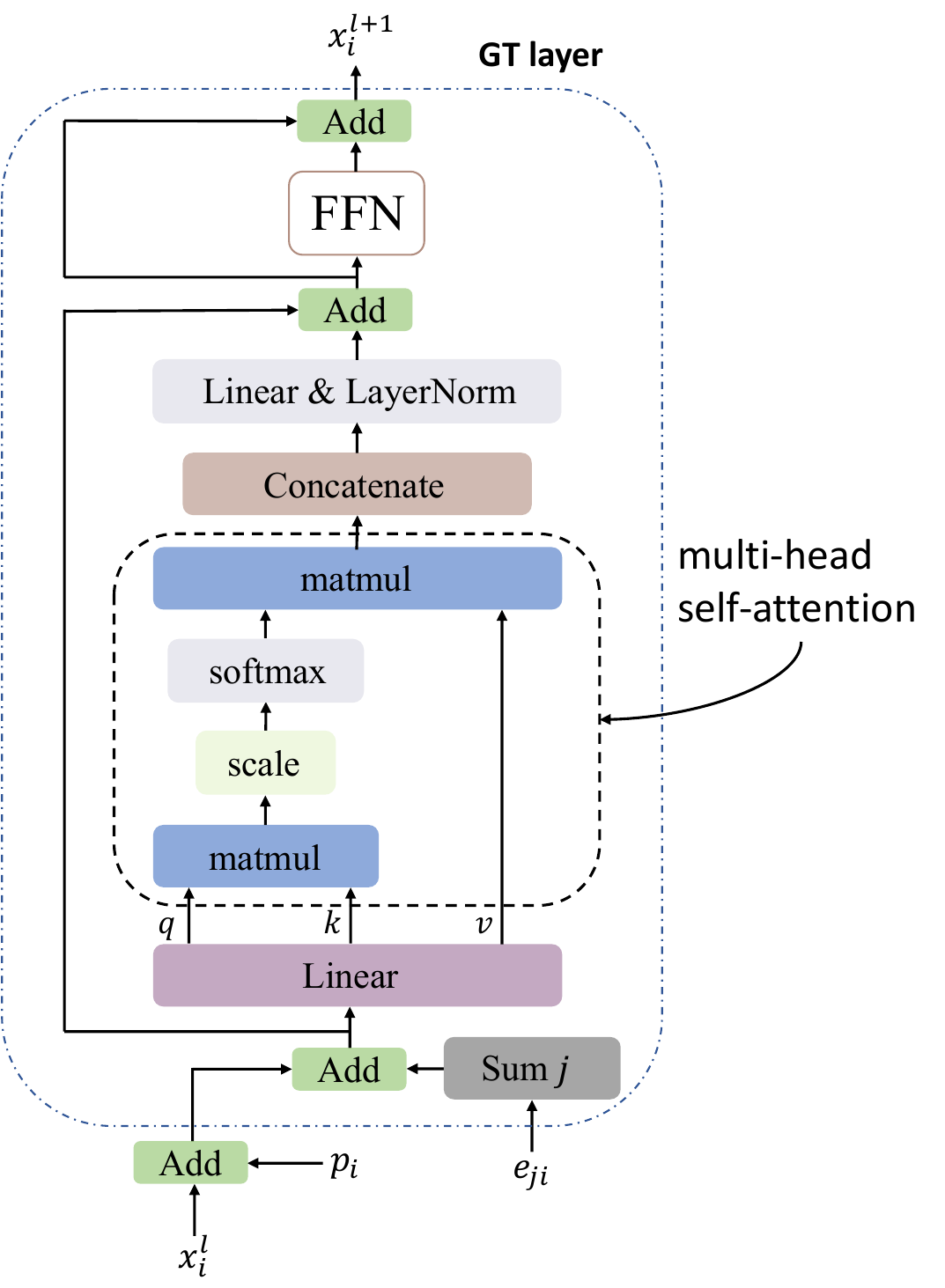}
    \caption{The architecture of a GT block. The GwT layer receives node embeddings ($x_i$), edge embeddings ($e_{ji}$), and positional encodings ($p_i$) as inputs, with only node embeddings undergoing iterative updates. Positional encodings ($p_i$) are directly summed with node embeddings ($x_i$) to explicitly integrate graph connectivity into node representations. Edge embeddings ($e_{ji}$) are aggregated via scattering and summation, then incorporated into node embeddings to encode structural relationships. The architecture adopts the classical transformer’s self-attention mechanism and feed-forward network (FFN), where node embeddings are dynamically refined using query ($q$), key ($k$), and value ($v$) projections. Central to the GwT layer is a graph-wise multi-head self-attention module, which establishes a fully connected framework to model long-range dependencies within the graph, extending transformer capabilities to non-sequential, graph-structured data.}
    \label{fig_ps:si_pred_gt}
\end{figure}

In the GT block, connectivity information is not explicitly included, as all nodes in each graph are considered an ordered sequence of tokens. Therefore, positional encoding and the sum of incoming edge features are combined with the node features beforehand:

\begin{equation}
x_{i,pe} = x_i + W_{pe}p_i + \sum_j W_ee_{ji} \label{eqg1}
\end{equation}
where $p_i$ denotes the positional encoding of node $i$.

A standard multi-head self-attention mechanism~\cite{vaswani2017attention} is utilized on the node features across the entire graphs, followed by a residual feed-forward network (FFN) that includes layer normalization:

\begin{equation}
  Q=W_qX, K=W_kX, V=W_vX \label{eqg2}\\
\end{equation}
\begin{equation}
  Attn=softmax(\frac{QK^T}{\sqrt{d_k}}) \label{eqg3}\\
\end{equation}
\begin{equation}
  H^l=X^{l}+LayerNorm(W_o(Attn\cdot V^{l})) \label{eqg4}\\
\end{equation}
\begin{equation}
  X^{l+1} = H^l + LayerNorm(W_2Gelu(W_1H^l+b_1)+b_2) \label{eqg5}
\end{equation}

The processed graphs are converted into graph representations using a mean pooling function applied to all node features. Following mean pooling, these graph representations are input into specialized feed-forward neural networks (FFNN) that predict the specific properties of the photosensitizer-solvent system. The FFNN can be denoted as:

\begin{equation}
    X^l = Softplus(W_lX^{l-1}+b_l) \label{eqf1} \\
\end{equation}
\begin{equation}
    X^{out} = W_nX^{n-1} + b_n \label{eqf2}
\end{equation}

where $X^{out}$ denotes the final prediction of the regression task. 

\subsection{Prediction Results on Photodynamic Properties}\label{si_pred_pred}

\begin{figure}
    \centering
    \includegraphics[width=0.9\linewidth]{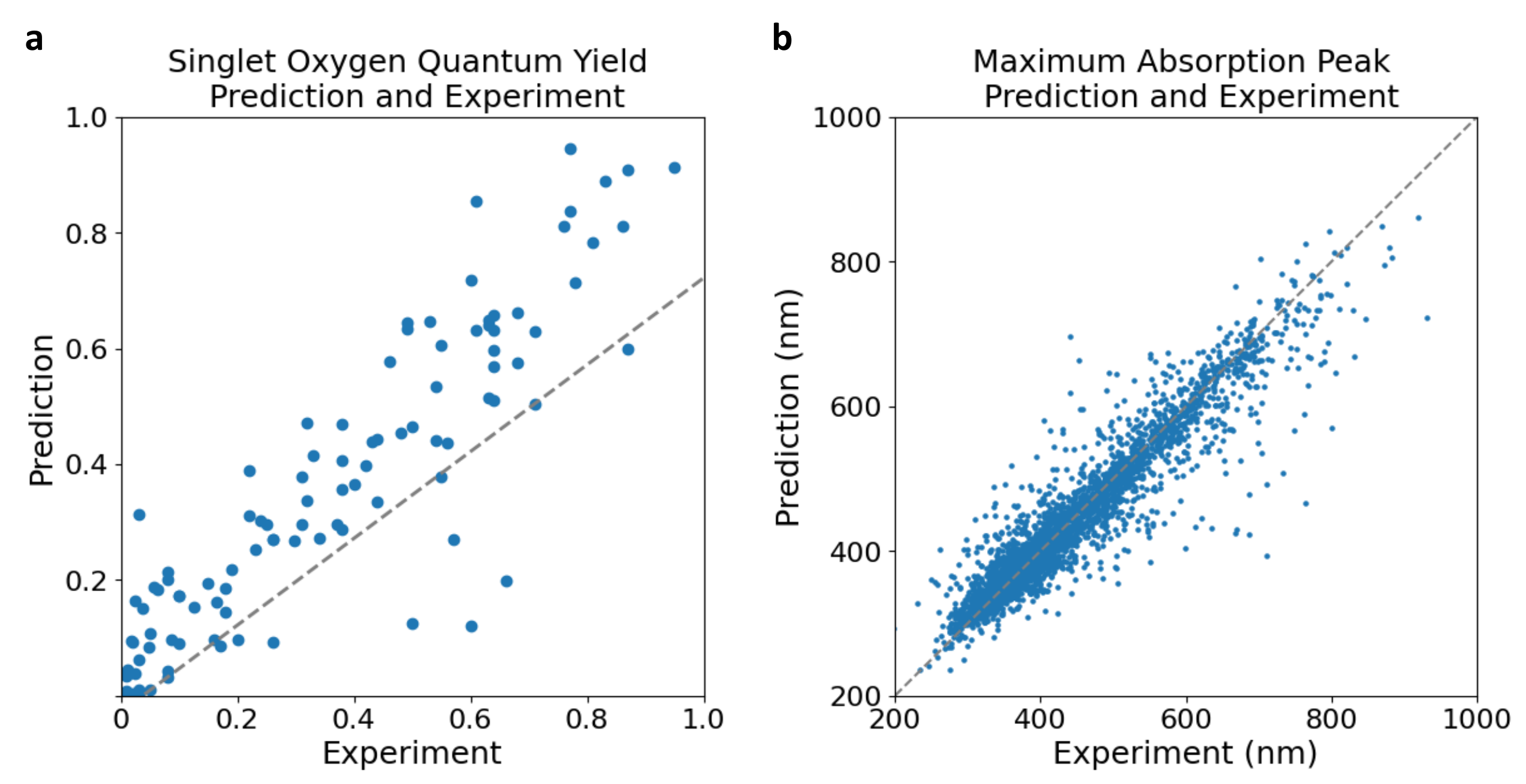}
    \caption{Correlation between SolutionNet predictions and experimental data for \textbf{a}: absorption maxima ($\lambda_{max}$) and \textbf{b}: singlet oxygen quantum yield ($\phi_\Delta$) across 5-fold cross-validation. Dashed lines represent the parity line ($y=x$).}
    \label{fig_ps:si_pred_corr}
\end{figure}

SolutionNet was evaluated for its ability to predict two critical photodynamic properties: absorption maxima ($\lambda_{max}$) and singlet oxygen quantum yield ($\phi_\Delta$). Using 5-fold cross-validation on experimental data, the model demonstrated robust performance, as illustrated in Figure~\ref{fig_ps:si_pred_corr}. For $\lambda_{max}$, predictions showed strong agreement with experimental values ($R^2$=0.888), reflecting the model’s capacity to link molecular structure to light absorption. The majority of predictions aligned closely with the parity line ($y=x$), with deviations primarily observed for molecules containing rare substituents or extended conjugation systems.

For $\phi_\Delta$, SolutionNet achieved a moderate correlation ($R^2$=0.704), underscoring the inherent complexity of modeling quantum yields, which depend on subtle electronic interactions and competing photophysical pathways. Despite this, the model effectively distinguished high-performance candidates ($\phi_\Delta$ $>$ 0.6) from low-yield molecules ($\phi_\Delta$ $<$ 0.3), enabling prioritization of promising candidates for experimental validation.

These results highlight SolutionNet’s utility as a screening tool for photosensitizer design, significantly reducing the experimental workload by identifying candidates with tailored photodynamic properties. The correlation plots in Figure~\ref{fig_ps:si_pred_corr} further validate its reliability in accelerating the discovery of next-generation photodynamic agents.

\section{Details of Molecule Generation}\label{si_gen}

\subsection{Architecture of the Generative Model}\label{si_gen_model}

The MoLeR~\cite{maziarz2021moler} model is a graph-based generative framework designed for scaffold-constrained molecule generation. Its architecture integrates graph neural networks (GNNs) and a motif-based vocabulary to efficiently construct novel molecules while preserving predefined core structures. Key components include:

\subsubsection{Graph Representation}
Molecules are represented as graphs where nodes correspond to atoms and edges encode bonds. This representation captures structural and chemical properties such as atom types and bond orders.

\subsubsection{Motif-Driven Generation}
MoLeR employs a predefined structural motif vocabulary, derived from decomposing training data into chemically meaningful fragments like functional groups. These motifs enable efficient generation of complex structures while avoiding unrealistic fragmentation.

\subsubsection{Encoder-Decoder Framework}

\textbf{Encoder}: A GNN maps the molecular graph into a latent vector, combining atomic features and motif embeddings to capture local and global patterns.

\textbf{Decoder}: Iteratively extends the scaffold by attaching motifs or atoms, guided by the latent vector. This process ensures generation is scaffold-centric and history-agnostic.

\subsubsection{Attention Mechanisms}
For dynamic scaffold-moiety alignment, MoLeR uses attention to prioritize motifs that align with the target scaffold’s chemical context, enhancing structural coherence.

This architecture balances efficiency and flexibility, enabling rapid generation of valid, diverse molecules while adhering to scaffold constraints.

% \subsection{Multi-objective Bayesian Optimization (MOBO)}\label{si_gen_mobo}

\subsection{Molecules at Pareto Frontier}\label{si_gen_frontier}

Molecules situated near the Pareto frontier in Figure~\ref{fig_ps:database}c are systematically evaluated for their synthetic accessibility and potential as PDT agents, guided by computational predictions and empirical expert assessment. Alongside the prioritized candidates (PNBD, HBS2N, HB4Ph), additional promising photosensitizers are presented in Figure~\ref{fig_ps:si_other_molecules}.

\begin{figure}
    \centering
    \caption{These molecules will be available after acceptance. Examples of 6 other photosensitizer candidates that we do not further investigate in this work. These molecules are generated from the Bayesian optimized generation and are predicted to be close to the Pareto frontier of long $\lambda_{max}$ and high $\phi_\Delta$. The 6 molecules includes derivatives of hypocrllin, hypericin and elsinochrome.}
    \label{fig_ps:si_other_molecules}
\end{figure}

\section{Details of TD-DFT Computation, Experiment and Characterization}\label{si_exp}

\subsection{Details of TD-DFT Computation}\label{si_exp_tddft}

\subsubsection{Computational Methodology}
All density functional theory (DFT) and time-dependent DFT (TD-DFT) calculations are performed using the Gaussian16 software package~\cite{frisch2016gaussian16}. The B3LYP hybrid functional~\cite{b3lyp}, combined with the 6-31G basis set~\cite{hehre1972631g,tirado2008631g_b3lyp}, is employed to optimize molecular geometries and compute electronic excitation properties. This approach balances computational efficiency and accuracy, particularly for organic systems, as B3LYP~\cite{b3lyp} reliably predicts ground-state geometries and excitation energies. TD-DFT calculations are conducted on the optimized ground-state (S$_0$ and T$_0$) structures to simulate absorption spectra and characterize excited-state transitions. The first few singlet and triplet excited states are evaluated to ensure comprehensive coverage of relevant electronic transitions.

The simulated absorption spectra for HB, HBS2N, PNBD, and HB4Ph (see Figure~\ref{si_exp_tddft}) reveal distinct absorption peaks corresponding to electronic transitions from the ground states to singlet (S$_1$) and triplet (T$_1$) excited states. The wavelength-dependent absorption behavior reflects variations in electronic structures.

\begin{figure}
    \centering
    \includegraphics[width=0.95\linewidth]{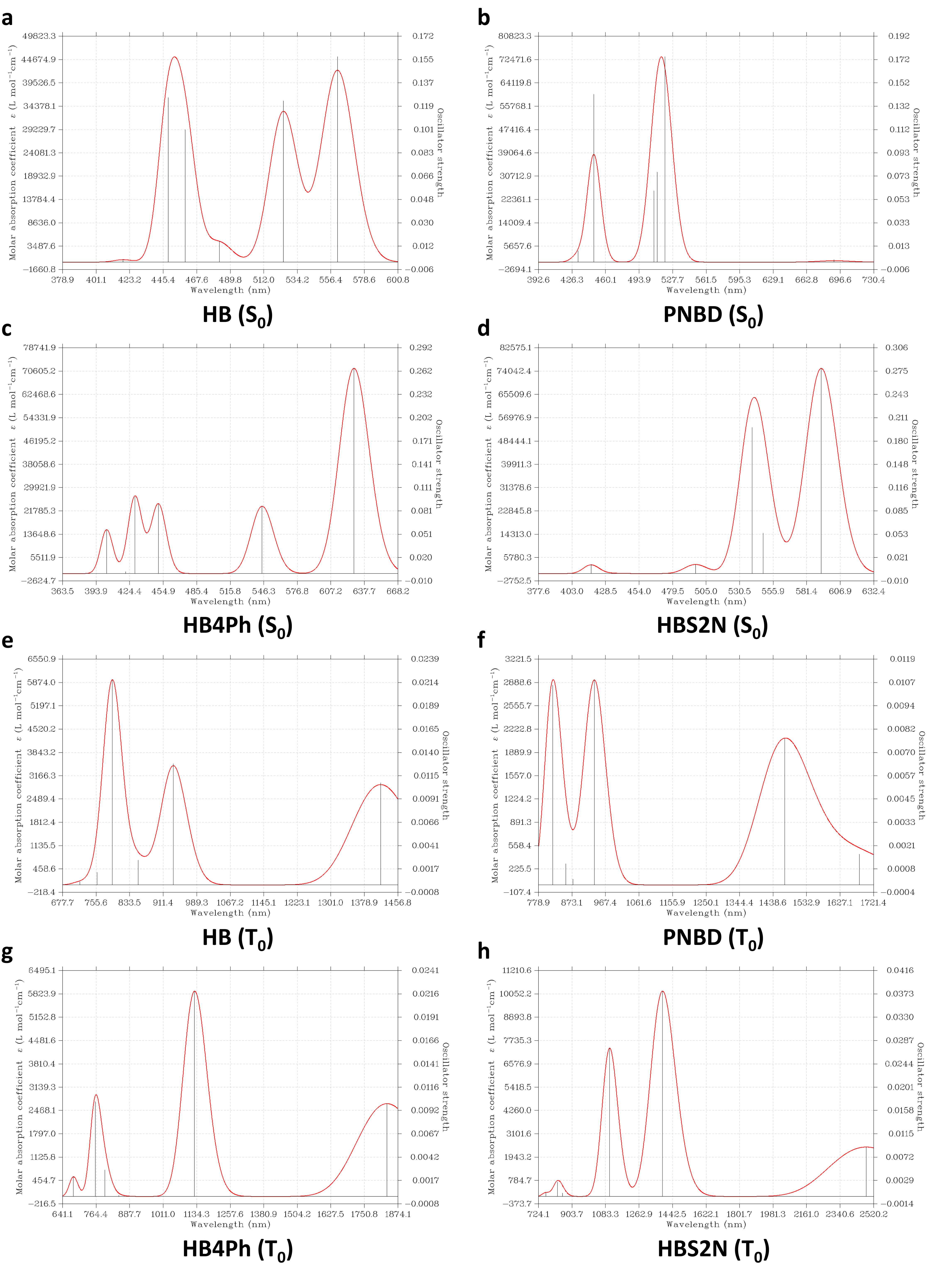}
    \caption{The absorption spectra predicted using TD-DFT computations. \textbf{a-d}: Absorption spectra at singlet ground state S$_0$ of HB (a), PNBD (b), HB4Ph (c) and HBS2N (d). \textbf{e-h}: Absorption spectra at triplet ground state T$_0$ of HB (a), PNBD (b), HB4Ph (c) and HBS2N (d). }
    \label{fig_ps:si_tddft_s0}
\end{figure}

\subsubsection{Energy Gap and Singlet Oxygen Prediction}

The energies of the first singlet (S$_1$) and triplet (T$_1$) excited states are derived from the computation results. The energy difference ($\Delta $E$_{st}$ = E(S$_1$) - E(T$_1$)) is calculated to qualitatively assess the potential efficiency of singlet oxygen ($^1$O$_2$) generation. Smaller $\Delta $E$_{st}$ values promote enhanced intersystem crossing (ISC) from S$_1$ to T$_1$, thereby increasing the likelihood of energy transfer to molecular oxygen and subsequent $^1$O$_2$ production. This computation provides insights into the PDT suitability of the studied compounds, with lower $\Delta $E$_{st}$ values correlating to higher predicted $^1$O$_2$ quantum yields. The computation results of $\Delta $E$_{st}$ can be found in Table~\ref{tab_ps:si_tddft_est}.

\begin{table}[]
 \caption{The computation results of $\Delta $E$_{st}$ of HB, PNBD, HB4Ph and HBS2N.}
    \centering
    \begin{tabular}{c|cccc}
        \hline
         &   HB& PNBD& HB4Ph&HBS2N\\
         \hline
         E$_{S1}$ (eV)&  2.2110& 1.7958& 1.9737&2.0930\\
         E$_{T1}$ (eV)& 0.8751& 0.7374& 0.6760&0.4999\\
         $\Delta$E$_{st}$ (eV)& 1.3359& 1.0584& 1.2977&1.5931\\
         \hline
     \end{tabular}
   
    \label{tab_ps:si_tddft_est}
\end{table}

% \begin{table}[]
%  \caption{The computation results of $\Delta $E$_{st}$ of HB, PNBD, HB4Ph and HBS2N.}
%     \centering
%     \begin{tabular}{c|cccc}
%         \hline
%          &   HB& PNBD& HB4Ph&HBS2N\\
%          \hline
%          E$_{S1}$ (hartree)&  -1834.0327& -2136.3837& -2102.7134&-2936.2974\\
%          E$_{T1}$ (hartree)& -1834.0460& -2136.2816& -2102.7208&-2936.3128\\
%          $\Delta$E$_{st}$ (hartree)& 0.0133& -0.0020& 0.0074&0.0155\\
%          $\Delta$E$_{st}$ (eV)& 0.3628& -0.0558& 0.2024&0.4210\\
%          \hline
%      \end{tabular}
   
%     \label{tab_ps:si_tddft_est}
% \end{table}  

\subsection{Details of Synthesis}\label{si_exp_synth}

% molecule figures, NMR
% \subsubsection{Intermediates}

\begin{figure}[H]
    \centering
    \includegraphics[width=0.9\linewidth]{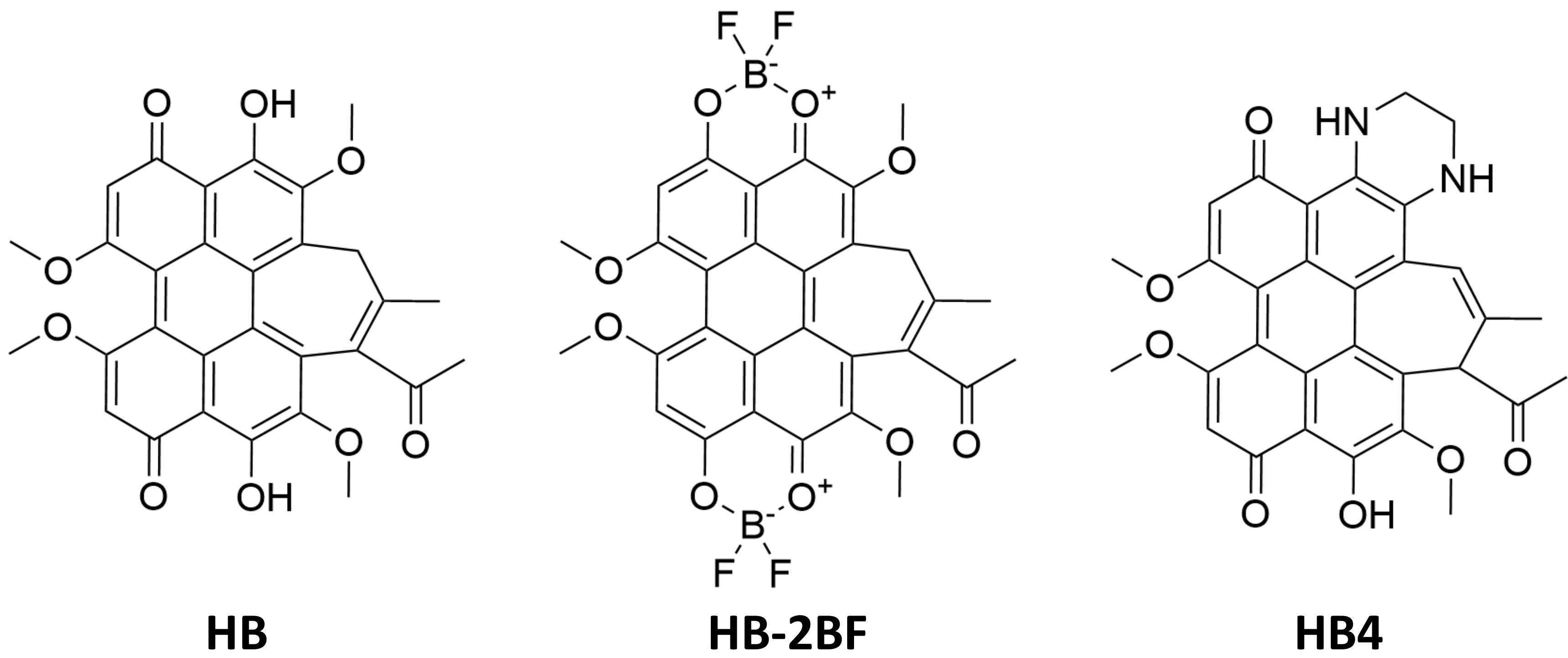}
    \caption{The molecular structure of precursors and intermediate products of the synthesis. HB serves as the precursor of the three photosensitizer candidates. HB-2BF is the intermediate product in the synthesis route of PNBD; HB4 is the intermediate product in the synthesis route of HB4Ph and HBS2N.}
    \label{fig_ps:si_intermediats}
\end{figure}

The synthesis of the three photosensitizer candidates is described in section~\ref{meth_synthesis_chara}. Here we show the intermediate products in Figure~\ref{fig_ps:si_intermediats}.

\begin{figure}
    \centering
    \includegraphics[width=0.95\linewidth]{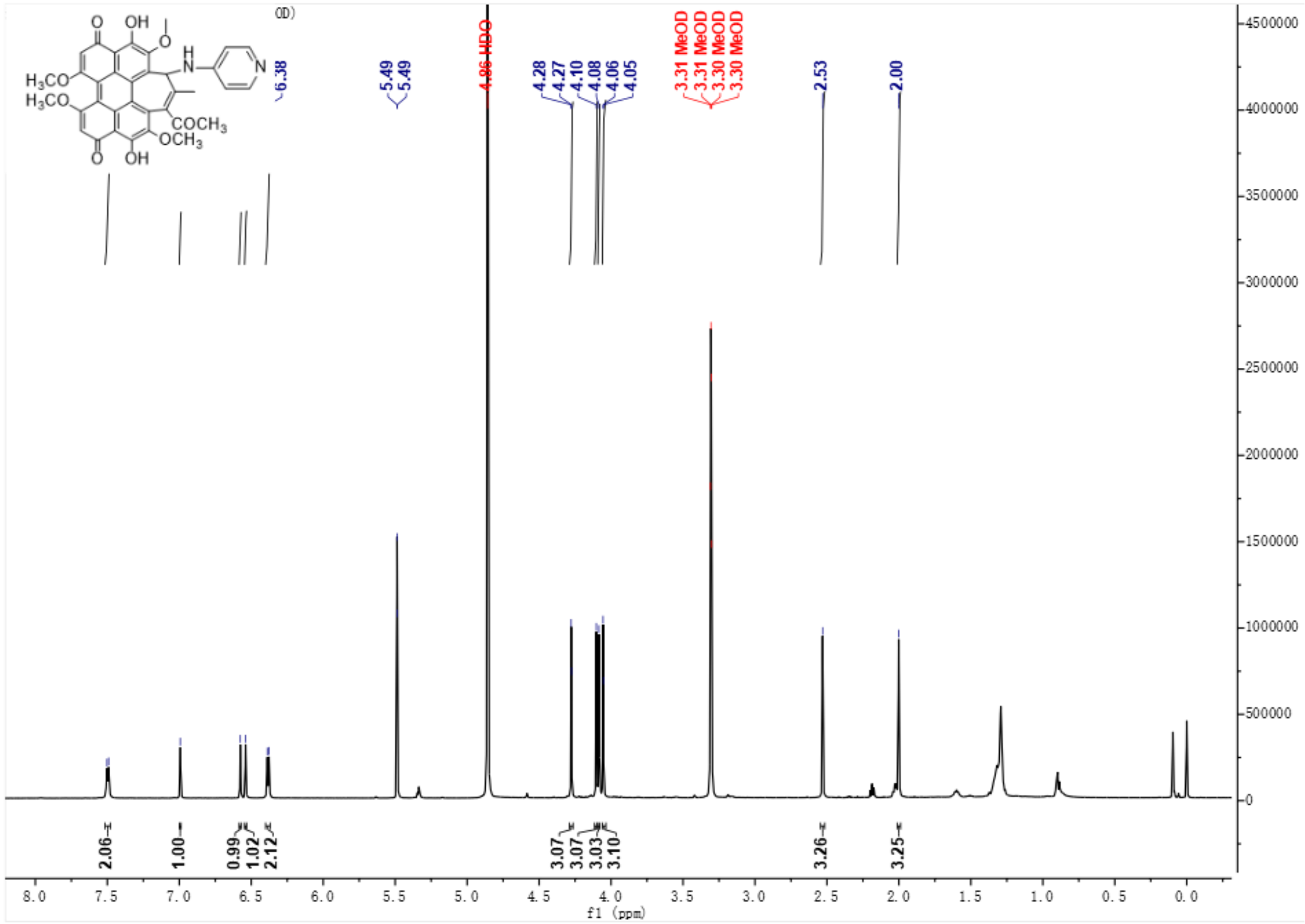}
    \caption{The $^1$H NMR spectrum of PNBD in Methanol-$d_4$.}
    \label{fig_ps:si_exp_nmr_pnbd}
\end{figure}

\begin{figure}
    \centering
    \includegraphics[width=0.95\linewidth]{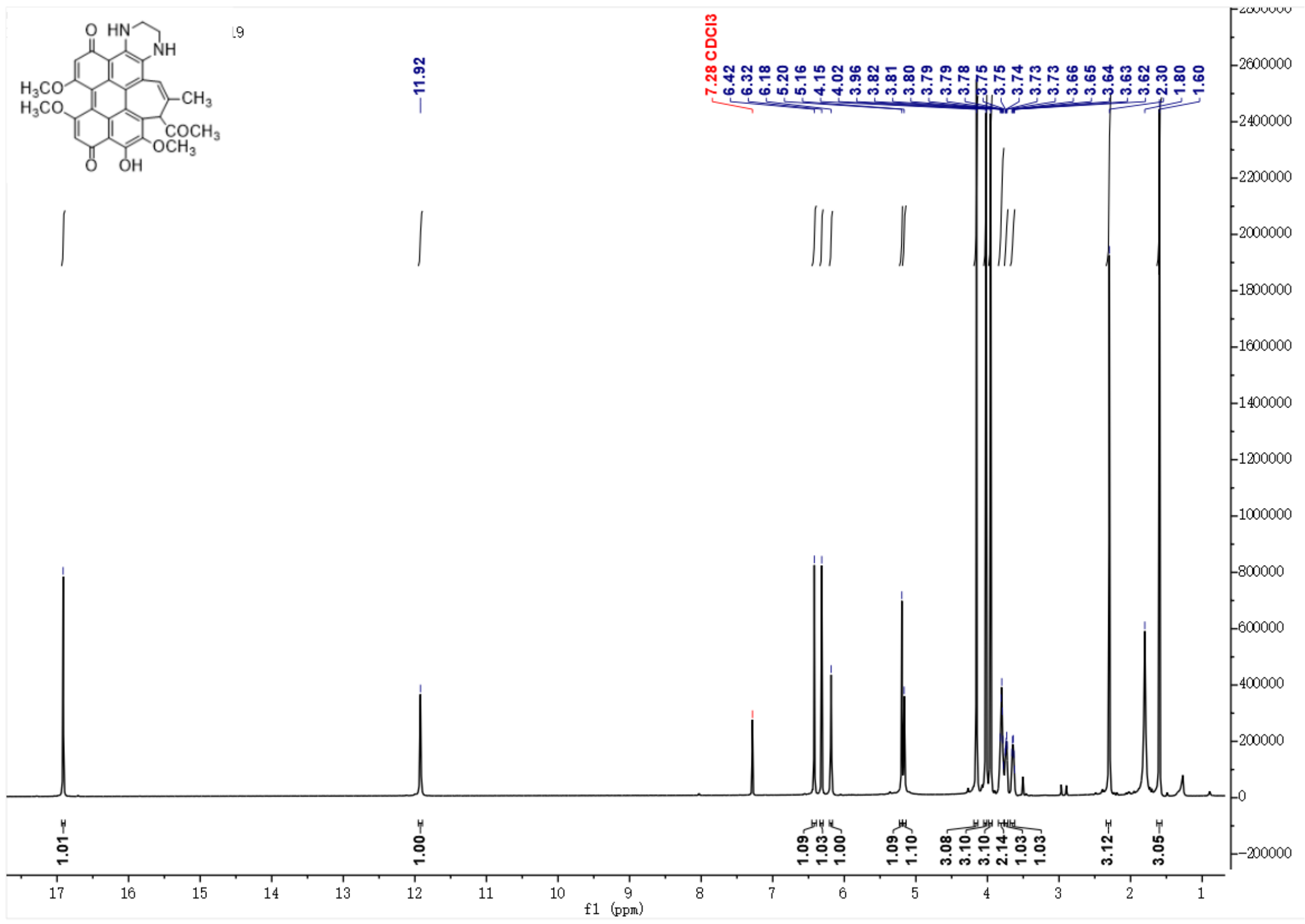}
    \caption{The $^1$H NMR spectrum of HB4 in Chloroform-$d$.}
    \label{fig_ps:si_exp_nmr_hb4}
\end{figure}

\begin{figure}
    \centering
    \includegraphics[width=0.95\linewidth]{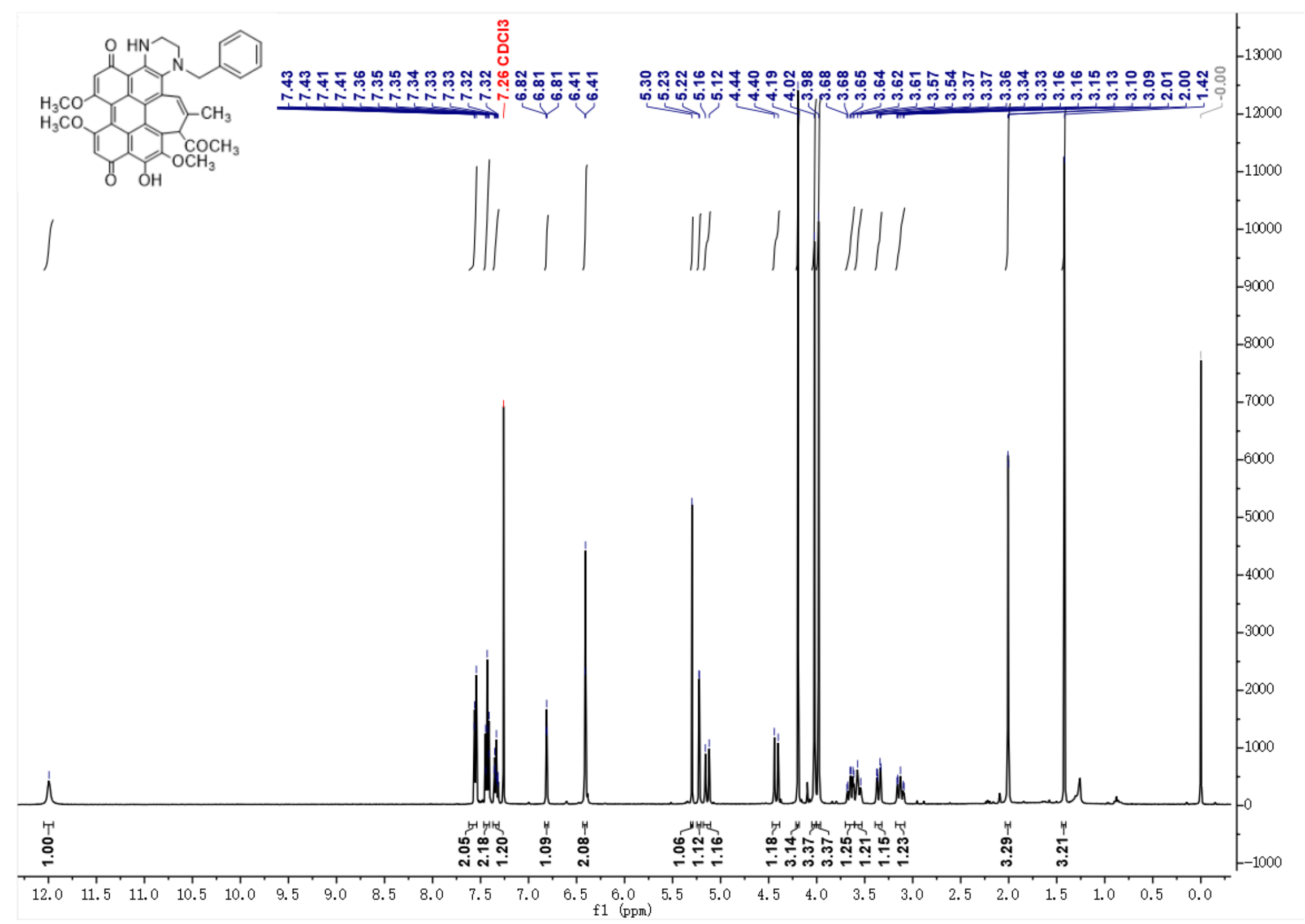}
    \caption{The $^1$H NMR spectrum of HB4Ph in Chloroform-$d$.}
    \label{fig_ps:si_exp_nmr_hb4ph}
\end{figure}

\begin{figure}
    \centering
    \includegraphics[width=0.95\linewidth]{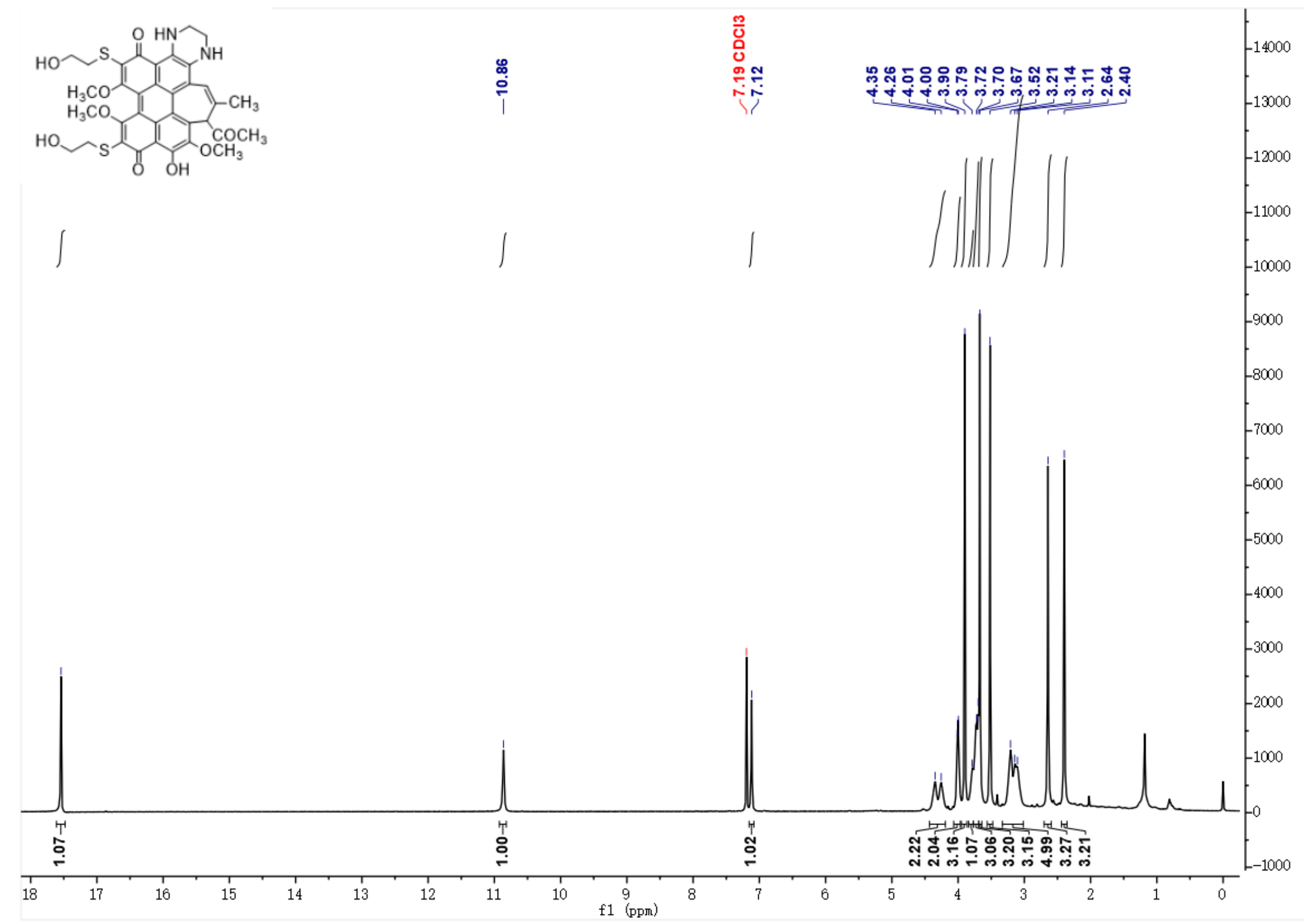}
    \caption{The $^1$H NMR spectrum of HBS2N in Chloroform-$d$.}
    \label{fig_ps:si_exp_nmr_hbs2n}
\end{figure}

After synthesis and purification of the intermediate product HB4 and the final products PNBD, HB4Ph and HBS2N, $^1$H-NMR spectra are tested to confirm the structure of the products. The $^1$H-NMR spectra are shown in Figure~\ref{fig_ps:si_exp_nmr_pnbd}-~\ref{fig_ps:si_exp_nmr_hbs2n}. % high-resolution mass spectra and 

% \begin{figure}
%     \centering
%     \includegraphics[width=0.9\linewidth]{}
%     \caption{Caption}
%     \label{fig_ps:si_exp_ms}
% \end{figure}

\subsection{Details of Characterization}\label{si_exp_charac}

\subsubsection{Absorption and Fluorescence Spectra detection}

The absorption and fluorescence spectra of HB, PNBD, HB4Ph, and HBS2N are characterized in dimethylformamide (DMF) solution. Absorption spectra are recorded using a UV-Vis spectrophotometer from 350 nm to 850 nm, with all spectra normalized to relative absorbance for direct comparison in a unified diagram. For fluorescence measurements, absorbance of each solution is calibrated to 0.2 at 530 nm prior to analysis, maintaining consistent excitation conditions across samples. Fluorescence spectra were acquired in a 1 cm $\times$ 1 cm quartz cuvette using a fluorescence spectrophotometer, with excitation at 530 nm and emission scanned from 550 to 850 nm.

\subsubsection{Singlet Oxygen Detection}\label{charac_so}

DPBF was employed as an indicator for $^1$O$_2$ detection, while HB acted as a standard reference. To rule out the influence of the inner-filter effect, the absorbance of HB, PNBD, HB4Ph, and HBS2N at 532 nm was calibrated to 0.2 OD in DMF. A sample solution (2 mL, prepared in DMF) was combined with 20 $\mu$L of DPBF (1 mg mL$^{-1}$, dissolved in DMF). Subsequently, the mixture's absorption spectra were monitored under 532 nm laser irradiation at 20 s intervals. The alteration in DPBF absorbance at 415 nm over a 120 s duration (normalized to its initial value as A/A$_0$) served as an indirect measure of the production rate of $^1$O$_2$.

\subsubsection{ROS Detection}\label{charac_ros}

DCFH was employed as the indicator for reactive oxygen species (ROS). Specifically, 1.980 mL of a DCFH aqueous solution (20 $\mu$M) was mixed with HB, PNBD, HB4Ph, or HBS2N (the absorbance of HB, PNBD, HB4Ph, and HBS2N at 532 nm was calibrated to 0.2 OD in DMF). Subsequently, the fluorescence spectra of DCFH were monitored under 532 nm laser irradiation at 10 s intervals. The increase in fluorescence intensity at 525 nm, represented as the ratio I/I$_0$, was utilized to evaluate the level of ROS generation.

\subsubsection{Spectra for Detection of Singlet Oxygen and ROS}

\begin{figure}
    \centering
    \includegraphics[width=0.95\linewidth]{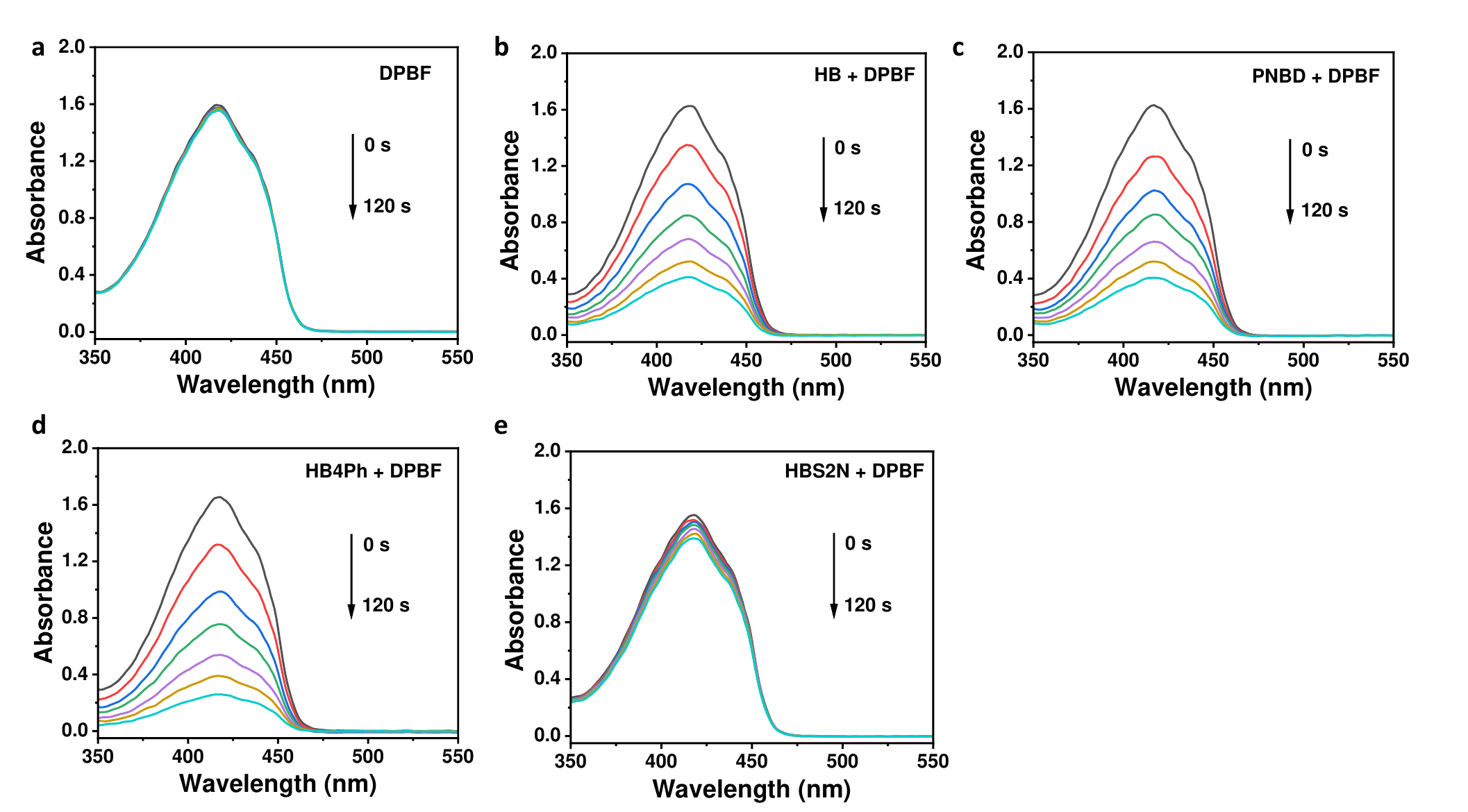}
    \caption{The detection of singlet oxygen with DPBF. \textbf{a}: DPBF solution without photosensitizer as a control group. \textbf{b, c, d, e}: Absorption spectra of DPBF and HB (b), PNBD (c), HB4Ph (d) and HBS2N (e) from 0 s to 120 s.}
    \label{fig_ps:si_exp_dpbf}
\end{figure}

\begin{figure}
    \centering
    \includegraphics[width=0.95\linewidth]{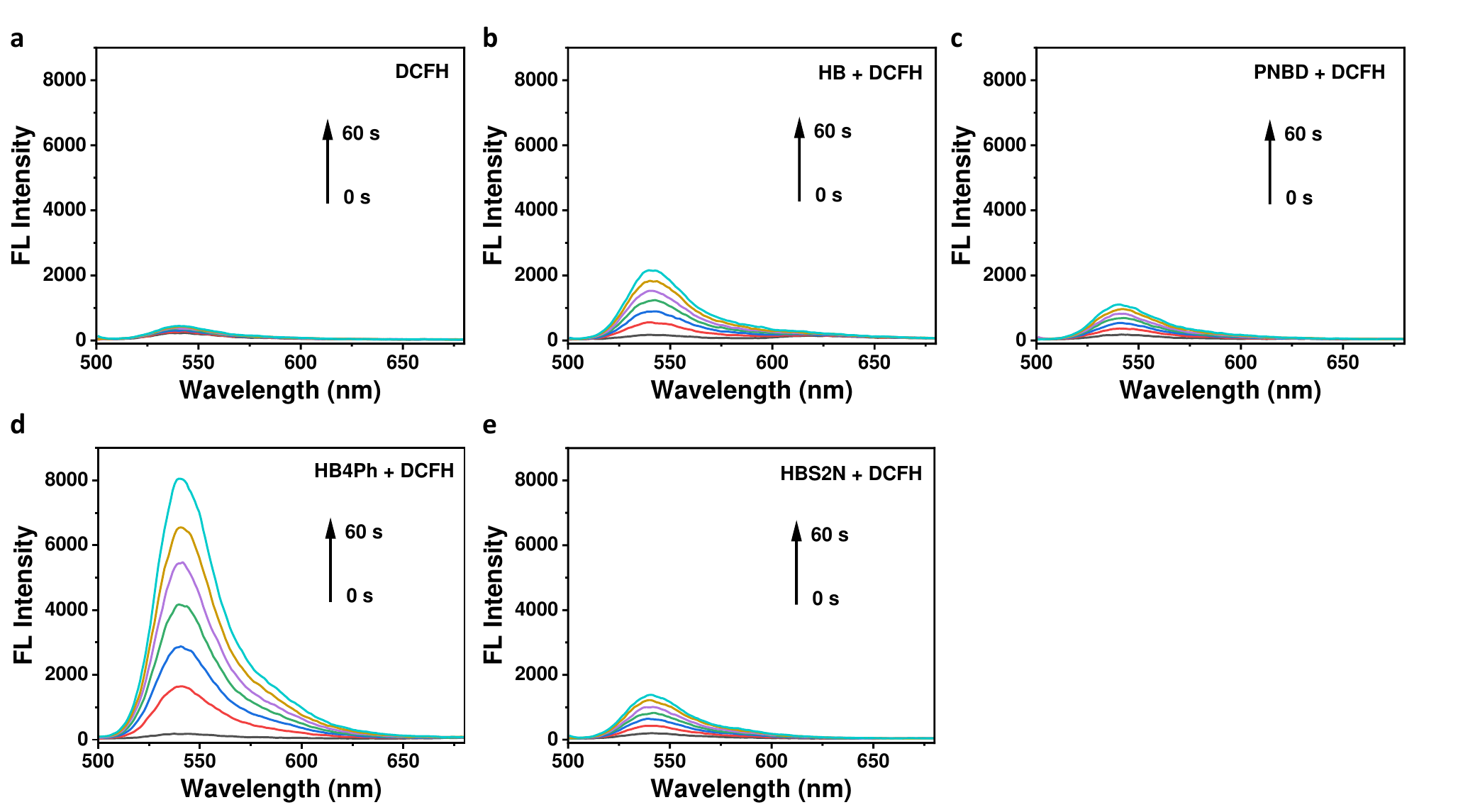}
    \caption{The detection of ROS with DCFH. \textbf{a}: DCFH solution without photosensitizer as a control group. \textbf{b, c, d, e}: Fluorescence spectra of DCFH and HB (b), PNBD (c), HB4Ph (d) and HBS2N (e) from 0 s to 60 s.}
    \label{fig_ps:si_exp_dcfh}
\end{figure}

The original absorption spectra of HB, PNBD, HBS2N, HB4Ph with DPBF for the detection of $\phi_\Delta$ is shown in Figure~\ref{fig_ps:si_exp_dpbf}, along with the control group of DPBF without photosensitizer. The original fluorescence spectra of the four photosensitizer candidates with DCFH for the detection of ROS generation is shown in Figure~\ref{fig_ps:si_exp_dcfh}, along with the control group of DCFH without photosensitizer.

\end{appendices}


%%\input latex-article.bbl

\end{document}